\documentclass[12pt]{article}
\usepackage[utf8]{inputenc}
\usepackage[british]{babel}
\usepackage{cmap}
\usepackage{lmodern}
\usepackage[T1]{fontenc}

\usepackage{amssymb, amsmath, amsthm}
\usepackage[a4paper,top=25mm,bottom=25mm,left=25mm,right=25mm]{geometry}
\usepackage{ragged2e}

\usepackage{authblk} 
\usepackage{pifont}
\usepackage{graphicx}
\usepackage[dvipsnames,svgnames,table]{xcolor}
\usepackage[figuresright]{rotating}
\usepackage{xtab} 
\usepackage{longtable} 
\usepackage{multirow}
\usepackage{footnote}
\usepackage[stable]{footmisc}
\usepackage{chngpage} 
\usepackage{pdflscape} 
\usepackage[nottoc,notlot,notlof]{tocbibind} 

\usepackage{pgfplots}
\pgfplotsset{every tick label/.append style={font=\footnotesize}}
\pgfplotsset{compat=1.18}
\usepackage{setspace}

\usepackage{array}
\newcolumntype{K}[1]{>{\centering\arraybackslash$}p{#1}<{$}}

\makesavenoteenv{tabular}
\usepackage{tabularx}
\usepackage{booktabs}
\usepackage{threeparttable}
\usepackage[referable]{threeparttablex} 
\newcolumntype{R}{>{\raggedleft\arraybackslash}X}
\newcolumntype{L}{>{\raggedright\arraybackslash}X}
\newcolumntype{C}{>{\centering\arraybackslash}X}
\newcolumntype{A}{>{\columncolor{gray!25}}C}
\newcolumntype{a}{>{\columncolor{gray!25}}c}

\newlength{\tablen}

\usepackage{dcolumn} 
\newcolumntype{.}{D{.}{.}{-1}}

\usepackage{tikz}
\usetikzlibrary{arrows, calc, matrix, patterns, positioning, trees}
\usepackage[semicolon]{natbib} 
\usepackage[hyphens]{xurl}
\usepackage{microtype}
\usepackage[justification=centering]{caption} 

\usepackage[labelformat=simple]{subcaption}

\DeclareCaptionLabelFormat{parenthesis}{(#2)}
\makeatletter
\renewcommand\p@subfigure{\arabic{figure}.}
\makeatother

\DeclareCaptionLabelFormat{parenthesis}{(#2)}
\makeatletter
\renewcommand\p@subtable{\arabic{table}.}
\makeatother

\usepackage{enumitem}

\setlist[itemize]{leftmargin=2.5\parindent}
\setlist[enumerate]{leftmargin=2.5\parindent}

\usepackage{hyperref} 
\hypersetup{
  colorlinks   = true,    		
  urlcolor     = blue,    		
  linkcolor    = blue,    		
  citecolor    = ForestGreen	
}

\def\addlegendimage{\csname pgfplots@addlegendimage\endcsname}

\theoremstyle{plain}

\theoremstyle{definition}

\theoremstyle{remark}

\makeatletter
\let\@fnsymbol\@alph
\makeatother

\def\keywords{\vspace{.5em} 
{\noindent \textit{Keywords}: }}

\def\AMS{\vspace{.5em} 
{\noindent \textbf{\emph{MSC} class}: }}

\def\JEL{\vspace{.5em} 
{\noindent \textbf{\emph{JEL} classification number}: }}

\title{A probabilistic match classification \\ model for low-scoring sports}

\author{\href{https://sites.google.com/view/laszlocsato}{L\'aszl\'o Csat\'o}\thanks{~Corresponding author. Email: \emph{laszlo.csato@sztaki.hun-ren.hu} \newline
Institute for Computer Science and Control (SZTAKI), Hungarian Research Network (HUN-REN), Laboratory on Engineering and Management Intelligence, Research Group of Operations Research and Decision Systems, Budapest, Hungary \newline
Corvinus University of Budapest (BCE), Institute of Operations and Decision Sciences, Department of Operations Research and Actuarial Sciences, Budapest, Hungary}
$\qquad \qquad$
Andr\'as Gyimesi\thanks{~Email: \emph{gyimesi.andras@ktk.pte.hu} \newline
University of P\'ecs, Faculty of Business and Economics, P\'ecs, Hungary \newline
Institute for Computer Science and Control (SZTAKI), Hungarian Research Network (HUN-REN), Laboratory on Engineering and Management Intelligence, Research Group of Operations Research and Decision Systems, Budapest, Hungary}}

\date{\today}

\def\Dedication{
{\noindent
``\emph{The pivotal change in the reforms announced by the UEFA Executive Committee is the departure from the current format's group stage system.
[\dots]
This gives the opportunity for clubs to test themselves against a wider range of opponents and raises the prospect for fans of seeing the top teams go head to head more often and earlier in the competition. It will also result in more competitive matches for every club across the board.}''
}
\vspace{0.25cm}

\flushright
\noindent
\small{(New format for Champions League post-2024: Everything you need to know \citep{UEFA2024c})}

\vspace{1cm} 
\justify }

\begin{document}
\newgeometry{top=25mm,bottom=25mm,left=25mm,right=25mm}
\maketitle

\thispagestyle{empty}
\Dedication

\begin{abstract}
\noindent
All existing match classification models in the tournament design literature suffer from two major limitations: a contestant is considered indifferent only if uncertain future results do never affect its prize, and competitive matches are not distinguished with respect to the incentives of the contestants. We propose a probabilistic framework to address both issues. For each match, our approach relies on simulating all other matches played simultaneously or later to compute the qualifying probabilities for the three main outcomes (win, draw, loss), thereby classifying each match into six categories. The suggested model is applied to the last round of the previous group stage and the new incomplete round-robin league, introduced in the 2024/25 season of UEFA club competitions. The incomplete round-robin tournament is found to contain fewer unimportant matches with two indifferent teams, and substantially more matches where both teams should play offensively. However, the robustly higher proportion of potentially collusive matches can threaten with serious scandals.


\AMS{62P20, 90-10, 90B90, 91B14}

\JEL{C44, C53, D71, Z20}
\end{abstract}

\clearpage
\restoregeometry

\section{Introduction} \label{Sec1}

The operational research literature has recently devoted serious effort to assess competitiveness in sports tournaments and to classifying the matches into different categories \citep{ChaterArrondelGayantLaslier2021, Csato2023a, CsatoMolontayPinter2024, Csato2025d, CsatoGyimesi2026a, DevriesereGoossensSpieksma2026, Gyimesi2024}.
Nonetheless, these models exhibit two major limitations.
First, previous papers always consider the incentives of the contestants as a binary concept: a contestant is \emph{indifferent} if and only if its prize is fully independent of the outcome of any match with an unknown result. However, it can be reasonably assumed that a probabilistic approach is more relevant in practice because the strategy of any contestant is likely not influenced by outcomes with a small chance of occurring. For example, a contestant may be satisfied with a particular result if it guarantees its qualification with a sufficiently high chance.

\begin{table}[t!]
\centering
\caption{Simulated qualifying probabilities in the last round \\ of the 2024/25 UEFA Conference League league phase}
\label{Table1}

\centerline{
\begin{threeparttable}
\rowcolors{3}{gray!20}{}
    \begin{tabularx}{1.1\linewidth}{ll CC CC CC} \toprule \hiderowcolors
    \multirow{2}[0]{*}{Home team} & \multirow{2}[0]{*}{Away team} & \multicolumn{2}{c}{Home win (1-0)} & \multicolumn{2}{c}{Draw (0-0)} & \multicolumn{2}{c}{Away win (0-1)} \\
          &       & Home & Away & Home & Away & Home & Away \\ \bottomrule \showrowcolors
    Borac Banja Luka  & Omonia  & 100\% & 29.48\% & 100\% & 97.97\% & 49.71\% & 100\% \\
    Molde & Mlad\'a Boleslav  & 98.41\% & 2.91\% & 0\%   & 93.67\% & 0\%   & 100\% \\ \toprule    
    \end{tabularx}
    \begin{tablenotes} \footnotesize
\item
\emph{Note}:
The column Home (Away) shows the probability that the home (away) team qualifies for the knockout stage play-offs based on 10 thousand simulations of the last round.
    \end{tablenotes}
\end{threeparttable}
}
\end{table}

Second, competitive matches, where none of the contestants is indifferent, can strongly differ depending on whether the incentives to play offensively or defensively are stronger.
As an illustration, consider Table~\ref{Table1}, which presents two matches played in the last matchday of the 2024/25 UEFA Conference League.
In the first match, both Borac Banja Luka and Omonia are expected to take few risks to score a goal: a draw of 0-0 ensures that both clubs qualify with a probability of more than 95\%, while even a small loss eliminates the loser in at least the half of all cases.
In the second match between Molde and Mlad\'a Boleslav, a draw is equivalent to a loss for Molde since none of them is sufficient for qualification, which creates strong incentives to score a goal. Contrarily, Mlad\'a Boleslav can be relatively satisfied with a goalless draw, which leads to an increase in the qualification probability by more than 90 percentage points compared to a small loss. Thus, the primary aim of Mlad\'a Boleslav is not to concede any goal.

\begin{table}[t!]
  \centering
  \caption{Basic categories of matches}
  \label{Table2}
\begin{threeparttable}
\rowcolors{1}{}{gray!20}
    \begin{tabularx}{\textwidth}{lCC} \toprule
    Match type & Incentives of Contestant A & Incentives of Contestant B \\ \bottomrule
    Unimportant & Indifferent & Indifferent \\
    Defensive asymmetric & Avoid losing & Indifferent \\
    Offensive asymmetric & Winning & Indifferent \\
    Antagonistic & Avoid losing & Winning \\
    Defensive & Avoid losing & Avoid losing \\
    Offensive & Winning & Winning \\ \bottomrule
    \end{tabularx}
\begin{tablenotes} \footnotesize
\item
\emph{Notes}: The two contestants are symmetric. Classification of incentives:
\item
Indifferent: strength of incentives to win $\sim$ strength of incentives to avoid losing
\item
Avoid losing: strength of incentives to win $\ll$ strength of incentives to avoid losing
\item
Winning: strength of incentives to win $\gg$ strength of incentives to avoid losing
\end{tablenotes}
\end{threeparttable}
\end{table}

In order to address these two issues, we present a general classification scheme for contests with two participants and three outcomes (this domain is called match for the sake of simplicity) in Section~\ref{Sec31}. For each match, our approach is based on simulating the results of all matches played simultaneously or later, which allows computing the probabilities of obtaining different prizes under the three main outcomes of the given match: a win for the first contestant, a draw, and a loss (win) for the first (second) contestant.
Each contestant can be
(1) indifferent; or it can have either
(2) stronger incentives to win than to avoid losing; or
(3) stronger incentives to avoid losing than to win.
This implies six distinct combinations as shown in Table~\ref{Table2}.

The proposed model is used to classify all matches played simultaneously in the last round of European club football competitions organised by the Union of European Football Associations (UEFA).
First, Section~\ref{Sec41} applies our approach to analyse the 2024/25 UEFA Conference League. Then, in Section~\ref{Sec42}, the relative frequency of the six categories of matches are computed for
(1) a home-away round-robin tournament played by four teams with eight representative distributions of team strengths (the 2023/24 UEFA Champions League groups);
(2) four incomplete round-robin tournaments with 36 teams and eight rounds (the 2024/25 and 2025/26 UEFA Champions League and UEFA Europa League);
(3) two incomplete round-robin tournaments with 36 teams and six rounds (the 2024/25 and 2025/26 UEFA Conference League).

We find that an incomplete round-robin tournament contains somewhat fewer unimportant, but substantially more defensive and offensive matches, especially with quite strong incentives. Unsurprisingly, playing six rounds instead of eight aggravates these effects, as well as reduces the proportion of asymmetric matches because the last round has a higher impact on the final ranking. While the incomplete round-robin format with 36 teams and six rounds indeed contains more competitive (antagonistic or offensive) matches than the traditional groups, the variant with eight rounds does not outperform double round-robin groups with four teams from this perspective---in contrast to the claim of UEFA, cited at the beginning of the paper. Nonetheless, replacing antagonistic matches with offensive matches may attract more spectators.

The proposed framework can be applied to any sport where draws are frequent and choosing an offensive option instead of a defensive one increases the probability of both winning and losing. These conditions certainly hold for association football, but might be relevant to chess, as well as to other relatively low-scoring sports such as field hockey and ice hockey, especially towards the end of a match.
In particular, while draws are prohibited in the National Hockey League (NHL), it would fit into our setting since a win in regulation time or overtime earns two points, a loss in overtime earns one point, and a loss in regulation time earns zero points. Thus, if goal difference is zero in the last minutes, teams should decide whether it is worth taking more risks for a higher chance of winning in regulation time, since this will also increase the probability of losing in regulation time.

\section{Literature review} \label{Sec2}

The related papers are discussed in three streams. Section~\ref{Sec21} gives an overview of the existing match classification schemes in order to highlight the novelty of our probabilistic approach. Section~\ref{Sec22} presents recent studies on incomplete round-robin tournaments, which are relevant to our application in Section~\ref{Sec4}. Finally, some empirical studies are reviewed in Section~\ref{Sec23} to demonstrate that football teams and fans react to the incentives created by the tournament. Consequently, match classification is not a mere operational research technique used in tournament design, but could have strong economic effects.

\subsection{Match classification models} \label{Sec21}

Match classification has been an extensively investigated topic in operational research over the last decade.
Inspired by the planned format of the 2026 FIFA World Cup, \citet{Guyon2020a} studies the risk of collusion in a single round-robin tournament with three teams when the two opponents in the last game know what results will let both of them advance. It is analysed how the risk of collusion depends on competitive balance and the match schedule. It is also shown that the high risk of collusion cannot be reduced by forbidding draws. These results have probably contributed to changing the format of the 2026 FIFA World Cup \citep{CsatoGyimesi2026a}.

\citet{Stronka2024} considers two innovative policies to mitigate the risk of collusion in this setting: randomised tie-breaking if some teams have the same number of points to generate extra uncertainty, and dynamic scheduling, which allows to determine the next match to be played based on the previous results. According to Monte Carlo simulations, the expected number of matches with a high risk of collusion decreases from 5.5 to 0.26 if both proposals are implemented.

Analysing groups of three teams is relatively simple since the matches are played sequentially. However, this condition does not hold for groups of four teams, where the last two matches are usually played simultaneously since the Disgrace of Gij\'on \citep[Section~3.9.1]{KendallLenten2017}.
\citet{ChaterArrondelGayantLaslier2021} present an assessment method for the competitiveness of the matches played in the last round in this case. Three types of matches are distinguished:
(a) competitive where neither team is indifferent and their targets are incompatible;
(b) collusive where neither team is indifferent and their targets are compatible (a non-empty subset of final scores exists that is optimal for both of them); and
(c) stakeless where at least one team is indifferent.
The frequencies of these matches are determined via simulations as a function of the schedule if two teams qualify from the group. Several alternative formats proposed for the FIFA World Cup with 48 teams are also studied.

\citet{Csato2023a} examines how tie-breaking rules affect the proportion of matches with at least one indifferent team in the four groups of the 2022/23 UEFA Nations League A, a double round-robin tournament with four teams. Giving priority to goal difference instead of head-to-head results can reduce the probability of a fixed position in the final group ranking by at least two percentage points in the last round.
\citet{Csato2025d} determines all collusion opportunities created by preferring head-to-head records for tie-breaking in a single round-robin tournament played by four teams. The probability of these situations is increased by at least 11 percentage points in the 2024 UEFA European Football Championship, which cannot be mitigated by any static match schedule.

\citet{CsatoMolontayPinter2024} introduce another match classification scheme with three categories:
(a) competitive if neither team is indifferent;
(b) weakly stakeless if exactly one team is indifferent; and
(c) strongly stakeless if both teams are indifferent.
The model is applied to a double round-robin contest with four teams, used in the previous group stage of UEFA club competitions. The probabilities of weakly and strongly stakeless matches are computed by a simulation model under the 12 valid schedules.
\citet{Gyimesi2024} calculates the proportion of these three types of matches for the pre-2024 and post-2024 designs of the UEFA Champions League, as well as the planned European Super League. The new incomplete round-robin league is found to reduce the number of stakeless matches played by the top teams.

\citet{CsatoGyimesi2026a} refine the concept of stakeless matches based on their expected outcome, as such a game is more costly if the indifferent team has a higher probability of winning \emph{a priori}. Monte Carlo simulations are used to estimate the probability that a given team becomes indifferent in the last round for two formats of the 2026 FIFA World Cup, the official format and an alternative containing imbalanced groups. The innovative design proposed greatly reduces the risk of a stakeless match played by the strongest teams.

Currently, the most detailed match classification scheme is given by \citet{DevriesereGoossensSpieksma2026}, which unifies the approaches of \citet{ChaterArrondelGayantLaslier2021} and \citet{CsatoMolontayPinter2024}. This implies four types of games: 
(a) asymmetric (equivalent to weakly stakeless in \citet{CsatoMolontayPinter2024});
(b) stakeless (equivalent to strongly stakeless in \citet{CsatoMolontayPinter2024});
(c) collusive;
(d) competitive if none of the above conditions holds.
Their probabilities are quantified via simulations for the pre-2024 and post-2024 formats of the UEFA Champions League. For the incomplete round-robin format, various schedules are generated based on reasonable ideas, such as pushing the most competitive matches to the first or to the last matchdays. The incomplete round-robin league is revealed to substantially reduce the proportion of asymmetric and stakeless matches under all schedules considered.


\subsection{Incomplete round-robin tournaments} \label{Sec22}

Research on incomplete round-robin tournaments has grown rapidly in recent years due to the introduction of incomplete round-robin leagues in UEFA club competitions from the 2024/25 season.
\citet{DevriesereGoossensSpieksma2026} and \citet{Gyimesi2024} have already been discussed in Section~\ref{Sec21}.
\citet{LiVanBulckGoossens2025} propose an approach using an incomplete round-robin format to solve the multi-league timetabling problem. Theoretical results are provided on the computational complexity of finding an incomplete round-robin tournament. The advantage of this novel format compared to the traditional multi-league round-robin is demonstrated via extensive experiments. One of the main benefits is reducing travel time, which persuaded the Belgian Royal Hockey Association to adopt this innovative design for its youth competitions \citep{DevriesereGoossens2025}.
\cite{deWerraUrrutiaAssuncao2026} analyse the problem of minimizing breaks (consecutive home or away games for a team) in incomplete round-robin schedules.

Three unpublished working papers examine the new design of UEFA club competitions.
\citet{CsatoGyimesiGoossensDevriesereLambersSpieksma2026} aim to measure the uncertainty of the draw in the UEFA Champions League by computing the variance of qualifying probabilities generated by the draw for each team.
\citet{GuyonBenSalemBuchholtzerTanre2026} compare four methods for drawing the league phase schedule, including the official draw procedure. Their fairness and impact on the matchup probabilities are studied.
\citet{WinkelmannDeutscherMichels2026} develop a bivariate Dixon--Coles simulation model \citep{DixonColes1997}, which adjusts the probabilities of low scoring games, to predict qualification thresholds in the UEFA Champions League and the UEFA Europa League.


\subsection{Empirical relevance of match classification} \label{Sec23}

One may ask whether the proposed quantification of incentives is supported by empirical evidence.
\citet{BuraimoForrestMcHaleTena2022} detect elevated interest for English Premier League matches with strong implications regarding the main prizes available for the clubs, such as the championship, UEFA Championship qualification, or relegation. For instance, a match with the highest championship significance in their dataset is expected to increase aggregate audience size by 96\%, while the match with the highest relegation significance attracts an audience size 54\% higher compared to a match with similar characteristics but no importance for seasonal outcomes.
According to \citet{FeddersenHumphreysSoebbing2023}, participants in betting markets believe that European football clubs adjust their effort level in line with the incentives provided by the structure of the league towards the end of the season. In particular, the expected reduced effort from indifferent teams is incorporated in the betting odds.
Analogously, teams played a draw significantly more frequently when it was mutually beneficial in historical FIFA World Cups and UEFA European Championships, and this draw benefit is priced into betting odds \citep{Medina2026}.


\section{Methodology} \label{Sec3}

Our match classification scheme attempts to resolve the two limitations mentioned in Section~\ref{Sec1}.
Therefore, for any given match, all other matches of the tournament played at the same time or later are simulated.
Next, we compare the increase in the probability of obtaining a prize by winning instead of playing a draw with the decrease in the probability of obtaining this prize by losing instead of playing a draw, in order to determine the potential gains and losses from playing offensively. The details are provided in Section~\ref{Sec31}.

The proposed match classification model will be applied to the last round of the pre-2024 and post-2024 formats of UEFA club competitions. Section~\ref{Sec32} presents the main characteristics of these tournaments.
The simulation method is described in Section~\ref{Sec33}, and the limitations of our model are discussed in Section~\ref{Sec34}.

\subsection{A match classification scheme based on incentives} \label{Sec31}

Let us take a match between two contestants, where the possible outcomes are $m$-$0$, $0$-$m$, and $0$-$0$. Even though an outcome $m$-$n$ is also possible in several sports, they are disregarded for the sake of simplicity. An offensive tactic is assumed to be associated with a higher probability of both scoring and conceding a goal.

Assume that the match is part of a tournament, where many contestants compete to earn some prizes. Denote the set of available prizes by $\mathcal{R}$. We define the possible gain $\mathcal{G}_i$ and possible loss $\mathcal{L}_i$ of contestant $i$ from playing more offensively as follows:
\[
\mathcal{G}_i = \max_{k \in \mathcal{R}} P_{i, k}^{(m\text{-}0)} - P_{i, k}^{(0\text{-}0)},
\]
\[
\mathcal{L}_i = P_{i, k}^{(0\text{-}0)} - P_{i, k}^{(0\text{-}m)},
\]
where $P_{i, k}^{(h)}$ is the probability that contestant $i$ obtains prize $k \in \mathcal{R}$ if the result of its match is $h$.
In other words, $\mathcal{G}_i$ indicates the gain from a win of $m$-$0$ compared to playing a draw of $0$-$0$, while $\mathcal{L}_i$ indicates the loss from a loss of $0$-$m$ compared to playing a draw of $0$-$0$. Naturally, both gains and losses are evaluated with respect to the probability of obtaining a prize specified by the rules of the tournament. Clearly, if $\mathcal{G}_i$ is relatively high (low) compared to $\mathcal{L}_i$, contestant $i$ has strong incentives to attack (defend).

The result $0$-$0$ is chosen as a benchmark since each match starts with this result. Thus, we quantify the incentives of the contestants at the beginning of the match. Naturally, these incentives change dynamically if a goal is scored either in the given match or in another match played simultaneously.

Since all probabilities are estimated from simulations, an indifference threshold $\mathcal{I}$ is introduced to avoid the possible distortion caused by a low-probability event: we say that contestant $i$ is indifferent if both $\mathcal{G}_i$ and $\mathcal{L}_i$ remain below the exogenously given parameter $\mathcal{I}$. 

As we have already seen in Table~\ref{Table2}, there are six distinct types of matches. In particular, the game played by contestants $i$ and $j$ is:
\begin{itemize}
\item
\emph{Unimportant} if $\max \left\{ \mathcal{G}_i; \mathcal{L}_i; \mathcal{G}_j; \mathcal{L}_j \right\} \leq \mathcal{I}$;

\item
\emph{Defensive asymmetric} if $\max \left\{ \mathcal{G}_i; \mathcal{L}_i \right\} \leq \mathcal{I}$ but $\max \left\{ \mathcal{G}_j; \mathcal{L}_j \right\} > \mathcal{I}$ and $\mathcal{G}_j < \mathcal{L}_j - \mathcal{I}$;

\item
\emph{Offensive asymmetric} if $\max \left\{ \mathcal{G}_i; \mathcal{L}_i \right\} \leq \mathcal{I}$ but $\max \left\{ \mathcal{G}_j; \mathcal{L}_j \right\} > \mathcal{I}$ and $\mathcal{G}_j \geq \mathcal{L}_j - \mathcal{I}$;

\item
\emph{Antagonistic} if $\max \left\{ \mathcal{G}_i; \mathcal{L}_i \right\} > \mathcal{I}$ and $\max \left\{ \mathcal{G}_j; \mathcal{L}_j \right\} > \mathcal{I}$ such that $\mathcal{G}_i < \mathcal{L}_i - \mathcal{I}$ and $\mathcal{G}_j \geq \mathcal{L}_j - \mathcal{I}$;

\item
\emph{Defensive} if $\max \left\{ \mathcal{G}_i; \mathcal{L}_i \right\} > \mathcal{I}$ and $\max \left\{ \mathcal{G}_j; \mathcal{L}_j \right\} > \mathcal{I}$ such that $\mathcal{G}_i < \mathcal{L}_i - \mathcal{I}$ and $\mathcal{G}_j < \mathcal{L}_j - \mathcal{I}$;

\item
\emph{Offensive} if $\max \left\{ \mathcal{G}_i; \mathcal{L}_i \right\} > \mathcal{I}$ and $\max \left\{ \mathcal{G}_j; \mathcal{L}_j \right\} > \mathcal{I}$ such that $\mathcal{G}_i \geq \mathcal{L}_i - \mathcal{I}$ and $\mathcal{G}_j \geq \mathcal{L}_j - \mathcal{I}$.
\end{itemize}
Hence, the contestants tend to play offensively by default, and choose a defensive strategy only if the payoff associated with avoiding a loss ($\mathcal{L}_i$) is higher than the payoff associated with winning ($\mathcal{L}_i$) by the ``margin of error'' $\mathcal{I}$.

In an unimportant game, both contestants are indifferent.
In a defensive asymmetric game, one contestant is indifferent, and the other contestant has stronger incentives to avoid a loss than to win.
In an offensive asymmetric game, one contestant is indifferent, and the other contestant has at least as strong incentives to win as to avoid a loss.

In an antagonistic game, neither contestant is indifferent; one contestant has stronger incentives to avoid a loss than to win, while the other contestant has at least as strong incentives to win as to avoid a loss.
In a defensive game, neither contestant is indifferent; and both contestants have stronger incentives to avoid a loss than to win.
In an offensive game, neither contestant is indifferent; and both contestants have at least as strong incentives to win as to avoid a loss.

A defensive game---although similar---does not coincide with a collusive game defined in the previous literature \citep{ChaterArrondelGayantLaslier2021, Csato2025d, DevriesereGoossensSpieksma2026}. In a collusive game, a particular outcome, either a draw or a win/loss with a small margin, guarantees the best possible prize for both contestants. Therefore, the match Borac Banja Luka vs Omonia, discussed in Section~\ref{Sec1}, is not collusive since Omonia is eliminated with a positive probability if the result is 0-0. But it is a defensive game, where tacit collusion is attractive, because the incentives to avoid losing are more powerful than to win for both contestants.

The difference between winning and avoiding a loss for team $i$ can be quantified as
\[
\omega_i =
\left\{
\begin{array}{ll}
0 & \text{if } \max \left\{ \mathcal{G}_i, \mathcal{L}_i \right\} \leq \mathcal{I}; \\
\lvert \mathcal{G}_i - \mathcal{L}_i \rvert & \text{if } \max \left\{ \mathcal{G}_i, \mathcal{L}_i \right\} > \mathcal{I}.
\end{array}
\right.
\]
If contestants $i$ and $j$ play against each other, the strength of incentives $\kappa$ in a(n)
\begin{itemize}
\item 
unimportant match is $\kappa = 0$;
\item 
defensive/offensive asymmetric match is $\kappa = 100 \times \max \{ \omega_i ; \omega_j \}$;
\item
antagonistic/defensive/offensive match is $\kappa = 100 \times \min \left\{ \omega_i ; \omega_j \right\}$.
\end{itemize}
Therefore, $\kappa$ is based on the incentives of the contestant that is not indifferent in an asymmetric match, and on the minimal incentives for the two contestants in an antagonistic/defensive/offensive match.
Note that $0 \leq \kappa \leq 100$.
In Section~\ref{Sec4}, we will call the incentives \emph{moderate} if $0 \leq \kappa < 25$, and \emph{strong} otherwise. $\kappa \geq 25$ means that the difference between the gain from winning and the loss from losing reaches 25\% in terms of the probability of obtaining the relevant prize for \emph{both} contestants.

\subsection{The tournament formats of UEFA club competitions} \label{Sec32}

In the 21 seasons played between 2003/04 and 2023/24, the most prestigious UEFA club competition, the UEFA Champions League, contained a group stage with eight groups of four teams each. A home-away round-robin tournament was played in each group; the first two teams qualified for the Round of 16, and the third-placed team was transferred to the second most prestigious UEFA club competition, called UEFA Cup until 2008/09 and UEFA Europa League after. The primary tie-breaking rules were head-to-head results, followed by goal difference and the number of goals scored.
Similar to \citet{DevriesereGoossensSpieksma2026}, the first prize is assumed to be the top two positions, and the second prize is assumed to be the third place in every group.


Since the 2024/25 season, the first stage of the three UEFA club competitions is a single incomplete round-robin league. The 36 teams play eight (UEFA Champions League and the UEFA Europa League) or six (UEFA Conference League) matches each, half of them at home and half of them away.
The top eight teams directly qualify for the Round of 16. The next 16 teams play against each other in the knockout stage play-offs, and the winners advance to the Round of 16. The ranking is based on the number of points scored, followed by goal difference and the number of goals scored \citep{CsatoDevriesereGoossensGyimesiLambersSpieksma2026}.
Following \citet{DevriesereGoossensSpieksma2026}, the first prize is the top eight positions, and the second prize is the positions from the 9th to the 24th.

This definition reduces the inherent bias caused by the prizes themselves in the comparison of the old and new designs of the UEFA Champions League: if all teams are of equal strength, they have a chance of 50\% and 22.2\% to obtain the first prize, as well as 75\% and 66.7\% to obtain at least the second prize, respectively. Even though the first prize is easier to obtain in the old format, teams also have a higher probability of avoiding elimination. Consequently, the change in the proportion of unimportant and asymmetric matches---where one or both teams are indifferent---caused by the reform is not predetermined by the prize structure of the pre-2024 and post-2024 designs.


\subsection{The simulation framework} \label{Sec33}

Simulating a sports tournament requires an assumption on the strength of the teams and a prediction model for match outcomes based on their strengths.
For the pre-2024 design, we simulate the eight groups of four teams each in the 2023/24 UEFA Champions League. In any group, there exist 12 different schedules with respect to the last round \citep{CsatoMolontayPinter2024}: a given team has three possible opponents and can play against them either at home or away, while the other match has two possible versions depending on which team plays at home.

For the post-2024 design, we consider the set of 36 clubs participating in the 2024/25 and 2025/26 UEFA Champions League, UEFA Europa League, and UEFA Conference League league phases. This provides six incomplete round-robin tournaments, four with eight rounds, and two with six rounds. In all these cases, the actual schedules are used since the principles of scheduling remain unknown, and it is far from trivial to generate further reasonable schedules \citep{DevriesereGoossensSpieksma2026, deWerraUrrutiaAssuncao2026, GuyonBenSalemBuchholtzerTanre2026}.

The 2024/25 and 2025/26 seasons of the UEFA Europa League are included in the analysis to increase the set of Elo ratings and schedules considered for the incomplete round-robin design. Since the Champions League and the Europa League share the same rules, the Europa League essentially provides a sensitivity analysis for the Champions League, given the limited historical data available for the Champions League.
The UEFA Conference League is worth considering because of the smaller number of rounds, six instead of eight. Thus, our simulations give an important insight into how the distribution of match types changes if fewer matches are played, which leads to a ``tighter'' final ranking and increases the impact of tie-breaking rules.

The strengths of the teams are measured by their Elo ratings. The Elo rating is updated after the team plays a match; hence, it changes throughout the season, but we follow the usual approach of fixing Elo ratings at the beginning of the season. In particular, Football Club Elo Ratings (\href{http://clubelo.com}{http://clubelo.com}) on 2 September, which is between the date of the league or group stage draw and the first match of the competition in every season, are used.

The standard choice for the distribution of the number of goals scored by a team is a Poisson distribution \citep{Maher1982, vanEetveldeLey2019}. We adopt this approach, analogous to the previous match classification literature \citep{ChaterArrondelGayantLaslier2021, CsatoGyimesi2026a, CsatoMolontayPinter2024, DevriesereGoossensSpieksma2026}. The expected number of goals scored in a match between teams $i$ and $j$ is assumed to be a cubic polynomial of win expectancy $W_{ij}$, which depends on the difference of the Elo ratings $E_i$ and $E_j$:
\begin{equation*}
W_{ij} = \frac{1}{1 + 10^{-(E_i - E_j)/400}}.
\end{equation*}
The cubic polynomials are estimated separately for the home and away team on the basis of all (almost 8000) matches played from 2003/04 to 2023/24 in UEFA club competitions and their qualifications. Further details and the exact values of the parameters can be found in \citet[Section~4.2]{CsatoGyimesiGoossensDevriesereLambersSpieksma2026}.

We aim to classify the matches played in the last round of the group stage and league phase when the results of previous rounds are known. Instead of using only the historical results of the competitions, which suffer from high randomness and do not provide an adequate sample size, all matches played before the last round are simulated 500 times for each competition design. Then, in each of these 500 scenarios, all matches played in the last round are simulated 1000 times. At first sight, it would be straightforward to compute the probability of obtaining a given prize from these 1000 simulations. However, this leads to only a limited number of wins (losses) for weak (strong) teams, making the frequencies observed in the simulations possibly inaccurate.

Therefore, since the classification scheme in Section~\ref{Sec31} requires comparing the probabilities of obtaining a prize under a win of $m$-$0$, a draw of $0$-$0$, and a loss of $0$-$m$, 1000 home wins of $m$-$0$ and 1000 away wins of $0$-$m$ are considered for each match such that the actual simulation run gives the results of all matches played simultaneously.
Fixing $m$ to a particular value would bias the probabilities (see Section~\ref{Sec41} for illustrations); thus, the probabilities of all $1 \leq m \leq 10$ are estimated for both the home and away teams in each match separately, using the match prediction model above.
For example, if the home team has a 30\% chance of scoring one goal, 35\% of scoring two goals, 10\% of scoring three goals, etc., then among the 1000 home wins, 300 are assumed to be $1$-$0$, 350 to be $2$-$0$, 100 to be $3$-$0$, etc. This ensures that the probability calculations are based on realistic goal differences. Finally, for each team, we calculate the probabilities of obtaining the prizes---top 2 and top 3 in the group stage, top 8 and top 24 in the league phase, as discussed in Section~\ref{Sec32}---under a win, a draw, or a loss.

Knowing the set of prizes $\mathcal{R}$, the probabilities $P_{m\text{-}0}^{(i, k)}$, $P_{0\text{-}0}^{(i, k)}$, $P_{0\text{-}m}^{(i, k)}$ directly determine the possible gain $\mathcal{G}_i$ and possible loss $\mathcal{L}_i$ for each team $i$.
The baseline value of the indifference threshold is $\mathcal{I}=0.02$, which has two roles. First, it controls the likelihood of unimportant and asymmetric matches, as will be studied in Section~\ref{Sec43} in more detail. Second, any team $i$ compares the possible gain $\mathcal{G}_i$ with $\mathcal{L}_i - \mathcal{I}$. This is important for the previous group format of the UEFA Champions League, where one team has exactly the same possible gain and possible loss in about 0.75\% of the simulations, which would make the classification of the match sensitive to whether the equality of implies $\mathcal{G}_i$ and $\mathcal{L}_i$ an offensive or defensive strategy. By comparing $\mathcal{G}_i$ to $\mathcal{L}_i - \mathcal{I}$, this uncertainty is fully eliminated.

For each group of the 2023/24 UEFA Champions League, 500 scenarios are created with each of the 12 possible match schedules. Since two matches are played in the last round, the probability of each match type is computed from a sample of 12000.
For the six incomplete round-robin leagues, only the actual schedule is used, but the last round consists of 18 matches, which results in a comparable sample size of 9000. Note that, although these numbers may seem small at first sight, the type of a match is also determined from the three outcomes for the given match (win of $m$-0 with a simulated $m$, draw of $0$-$0$, loss of 0-$m$ with a simulated $m$) and 1000 simulations of all matches played simultaneously, respectively.

\subsection{Limitations of the proposed probabilistic model} \label{Sec34}

Naturally, our approach has various limitations, similar to all models of the real-world.
First, the incentives of the teams are assumed to be static; they are computed only for the initial score of 0-0. In practice, the incentives evolve during the game if the result of the given match, or any match played simultaneously, changes. The incentives are more likely to change with a larger number of simultaneous matches. This rather affects the novel incomplete round-robin format, where 17 other matches are played at the same time, compared to the only one in the previous groups of four. Obviously, the simultaneous matches are expected to have a stronger effect on the incentives when
(1) fewer rounds are played (the assumption of static incentives is less realistic in the UEFA Conference League than in the UEFA Champions League and the UEFA Europa League);
(2) the competition is more balanced.

Nonetheless, we have two arguments for using static incentives. Making the incentives dynamic would require not only a model for the number of goals scored by the two teams, but a model of exact goal arrival times. While these models exist in the literature \citep{DixonRobinson1998, LeriouNtzoufras2025}, they do not have a well-established methodology at the moment.
In addition, static incentives are relevant when the goal is to select games in advance. For example, a fan could watch only one match in person, and a television company might want to broadcast live fewer than 18 simultaneously played matches.

Second, the teams are assumed to be na\"ive since they do not take the implied changes in the behaviour of \emph{other} teams into account when deciding about their strategy in the last match. This corresponds to bounded rationality, level-1 type in behavioural game theory: players believe that other players are level-0 types, that is, they do not think strategically \citep{StahlWilson1995}. However, the impact of incentives on the match results is currently unknown and cannot be incorporated into the simulation model without arbitrary assumptions. Hopefully, our first attempt to quantify incentives in a probabilistic framework will be followed by empirical studies exploring how these incentives change the dynamics of matches.

Third, the quantitative results depend on the assumed Poisson goal-generating process and the particular values of the Elo ratings. The independent Poisson model is the standard choice in the literature and is usually find to be competitive or even the best based on predictive accuracy \citep{ChaterArrondelGayantLaslier2021, LeyvandeWieleVanEeetvelde2019}. Analogously, the results are unlikely to be driven by the choice of the particular historical Elo ratings. Figure~\ref{Fig2} in Section~\ref{Sec42} will demonstrate that the 2023/24 UEFA Champions League groups cover a wide range of competitive settings in groups of four teams, and a sensitivity analysis is conducted in Section~\ref{Sec43} with respect to group composition. For the incomplete round-robin format, we use data from all historical tournaments, and this format is less sensitive to the draw than the previous group stage \citep{CsatoGyimesiGoossensDevriesereLambersSpieksma2026}.
Finally, even if the simulation model would be biased, it probably distorts the main characteristics of two tournament designs to the same extent.

Fourth, a model estimated on the basis of the previous format of UEFA club competitions can be invalid in the novel league phase due to the shifts in the incentives of the teams \citep{Csato2025h, WinkelmannDeutscherMichels2026}. Still, for comparability, we opt to use the same model in the case of the novel format, too.

Fifth, the definition of the available prizes is simplified.
In the Round of 16 until the 2023/24 season, the group winners played against the runners-up with hosting the second leg, which might have created some incentives to win the group. In the new design, the knockout brackets are almost fixed; for example, the team ranked 1st (8th) in the league phase plays against the winner of the playoff between the teams ranked the 15/16th and 17/18th (9/10th and 23rd/24th). The top two teams are also guaranteed to play the second leg at home in the Round of 16, quarterfinals, and semifinals since the 2025/26 season. Consequently, there are differences even among the top eight and the following 16 teams, which might create secondary incentives to win.

UEFA provides financial incentives to win more games \citep{Csato2023f}. Nonetheless, this prize money is only a small proportion of total revenue for clubs from the top European leagues due to their lucrative broadcast contracts \citep{Rewilak2026}.
Furthermore, it is far from clear that expected opponent quality is substantially more favourable if a team finishes in a better position in the group or the incomplete round-robin league \citep{Csato2025b, Guyon2022a}. As an illustration, both the 2024/25 and 2025/26 UEFA Champions League were won by Paris Saint-Germain, which finished only in the 15th and 11th position in the league phase, respectively.
Based on these arguments, we see no compelling reason to depart from the current definition of the prize structure.

\section{Applications} \label{Sec4}

Section~\ref{Sec41} focuses on the league phase of the 2024/25 UEFA Conference League, where the last round provides excellent illustrations of all categories of matches. This is followed by computing the probability of each match type in the eight groups of the 2023/24 UEFA Champions League, as well as in the six finished seasons of UEFA club competitions organised in an incomplete round-robin format. Finally, Section~\ref{Sec43} presents a sensitivity analysis with respect to the indifference threshold $\mathcal{I}$ and group balance in order to highlight the unique features of the incomplete round-robin format.

While we focus on the last round, where all matches are played simultaneously, teams may become indifferent even before the last round \citep{CsatoMolontayPinter2024}, and collusive matches can also occur in previous rounds \citep{DevriesereGoossensSpieksma2026}. The methodology is applicable to earlier rounds, but match classification has particular relevance in the last round, and the analysis of the previous rounds would be complicated by the order in which the matches are played.

\subsection{The last round of the 2024/25 UEFA Conference League} \label{Sec41}


The last matchday of the 2024/25 UEFA Conference League league phase was 19 December 2024. Table~\ref{Table_A1} in the Appendix shows the standing on the previous day. Recall from Section~\ref{Sec32} that the top eight clubs directly qualify for the Round of 16, the next 16 clubs advance to the knockout phase play-offs, and the last 12 are eliminated.

\begin{table}[t!]
\centering
\caption{Matches in the last round of the 2024/25 UEFA Conference League league phase}
\label{Table3}

\centerline{
\begin{threeparttable}
\rowcolors{3}{}{gray!20}
    \begin{tabularx}{1.12\linewidth}{ll Ccl} \toprule \hiderowcolors
    Home team & Away team & Result & Match type & Incentives $\kappa$ \\ \bottomrule \showrowcolors       
    \emph{1.~FC Heidenheim}  & St.~Gallen  & 1-1   & Offensive & 21.8 (Moderate) \\
    \emph{APOEL}  & Astana & 1-1   & Offensive & 67.1 (Strong) \\
    Borac Banja Luka  & \emph{Omonia}  & 0-0   & Defensive & 54.1 (Strong) \\
    \emph{Celje}  & The New Saints  & 3-2   & Offensive & 13.7 (Moderate) \\
    Cercle Brugge & \emph{{\. I}stanbul Ba{\c s}ak{\c s}ehir}  & 1-1   & Offensive & 71.2 (Strong) \\
    Chelsea  & \emph{Shamrock Rovers}  & 5-1   & Defensive asymmetric & 28.4 (Strong) \\
    \emph{Djurg{\r a}rdens IF}  & Legia Warsaw  & 3-1   & Antagonistic & 21.2 (Moderate) \\
    \emph{Heart of Midlothian}  & Petrocub H{\^ i}nce{\c s}ti  & 2-2   & Defensive asymmetric & 50.5 (Strong) \\
    \emph{Jagiellonia Bia{\l}ystok}  & Olimpija Ljubljana  & 0-0   & Offensive & 40.7 (Strong) \\
    Larne  & \emph{Gent}  & 1-0   & Offensive asymmetric & 46.2 (Strong) \\
    LASK  & \emph{V\'ikingur Reykjav\'ik}  & 1-1   & Defensive asymmetric & 23.1 (Moderate) \\
    Lugano  & \emph{Pafos}  & 2-2   & Antagonistic & 67.2 (Strong) \\
    \emph{Molde}  & Mlad\'a Boleslav  & 4-3   & Antagonistic & 85.4 (Strong) \\
    \emph{Panathinaikos}  & Dinamo Minsk  & 4-0   & Antagonistic & \textcolor{white}{0}2.6 (Moderate) \\
    \emph{Rapid Wien}  & Copenhagen  & 3-0   & Offensive & \textcolor{gray!20}{0}2.6 (Moderate) \\
    Real Betis  & \emph{HJK}   & 1-0   & Antagonistic & \textcolor{white}{0}4.0 (Moderate) \\
    \emph{TSC}   & Noah  & 4-3   & Offensive & 29.6 (Strong) \\
    Vit\'oria de Guimar{\~ a}es  & \emph{Fiorentina}  & 1-1   & Defensive asymmetric & \textcolor{white}{0}4.5 (Moderate) \\ \bottomrule    
    \end{tabularx}
\begin{tablenotes} \footnotesize
\item
\emph{Notes}:
The team with stronger incentives is written in \emph{italic} font.
\item
The last column shows the strength of incentives $\kappa$ based on the expected value of parameter $m$.
\end{tablenotes}
\end{threeparttable}
}
\end{table}

Table~\ref{Table3} classifies the 18 matches played simultaneously in the last round according to the scheme presented in Section~\ref{Sec31}, together with their final result.

\begin{figure}[t!]

\begin{tikzpicture}
\begin{axis}[
name = axis1,
title = {Offensive match: \\ APOEL vs Astana},
title style = {font=\small,align=center},
xlabel = Value of $m$ (goal difference),
x label style = {font=\small},
x tick label style={/pgf/number format/1000 sep=},
xtick = data,
ylabel = Change in probability,
width = 0.5\textwidth,
height = 0.38\textwidth,
ymajorgrids = true,
xmin = -5.2,
xmax = 5.2,
ymin = -1.05,
ymax = 1.05,
] 
\addplot [red, thick, dashed, mark=o, mark size=2pt, mark options=solid] coordinates {
(-5,-0.0016)
(-4,-0.0016)
(-3,-0.0016)
(-2,-0.0016)
(-1,-0.0016)
(0,0)
(1,0.9575)
(2,0.9888)
(3,0.9948)
(4,0.9977)
(5,0.9982)
};
\addplot [red, very thick, dotted, mark=none] coordinates {
(-5,-0.0016)
(-4,-0.0016)
(-3,-0.0016)
(-2,-0.0016)
(-1,-0.0016)
(0,0)
(1,0.9788)
(2,0.9788)
(3,0.9788)
(4,0.9788)
(5,0.9788)
};
\addplot [blue, thick, dashdotted, mark=asterisk, mark size=2.5pt, mark options={solid,semithick}] coordinates {
(-5,0.9928)
(-4,0.957)
(-3,0.9134)
(-2,0.8054)
(-1,0.5524)
(0,0)
(1,0)
(2,0)
(3,0)
(4,0)
(5,0)
};
\addplot [blue, very thick, dotted, mark=none] coordinates {
(-5,0.6706)
(-4,0.6706)
(-3,0.6706)
(-2,0.6706)
(-1,0.6706)
(0,0)
(1,0)
(2,0)
(3,0)
(4,0)
(5,0)
};
\end{axis}

\begin{axis}[
at = {(axis1.south east)},
xshift = 0.08\textwidth,
title = {Defensive match: \\ Borac Banja Luka vs Omonia},
title style = {font=\small,align=center},
xlabel = Value of $m$ (goal difference),
x label style = {font=\small},
x tick label style={/pgf/number format/1000 sep=},
xtick = data,
width = 0.5\textwidth,
height = 0.38\textwidth,
ymajorgrids = true,
xmin = -5.2,
xmax = 5.2,
ymin = -1.05,
ymax = 1.05,
] 
\addplot [red, thick, dashed, mark=o, mark size=2pt, mark options=solid] coordinates {
(-5,-0.6848)
(-4,-0.6293)
(-3,-0.5666)
(-2,-0.5394)
(-1,-0.5029)
(0,0)
(1,0)
(2,0)
(3,0)
(4,0)
(5,0)
};
\addplot [red, very thick, dotted, mark=none] coordinates {
(-5,-0.541)
(-4,-0.541)
(-3,-0.541)
(-2,-0.541)
(-1,-0.541)
(0,0)
(1,0)
(2,0)
(3,0)
(4,0)
(5,0)
};
\addplot [blue, thick, dashdotted, mark=asterisk, mark size=2.5pt, mark options={solid,semithick}] coordinates {
(-5,0.0203)
(-4,0.0203)
(-3,0.0203)
(-2,0.0203)
(-1,0.0203)
(0,0)
(1,-0.6849)
(2,-0.6877)
(3,-0.7717)
(4,-0.8379)
(5,-0.8924)
};
\addplot [blue, very thick, dotted, mark=none] coordinates {
(-5,0.0203)
(-4,0.0203)
(-3,0.0203)
(-2,0.0203)
(-1,0.0203)
(0,0)
(1,-0.6981)
(2,-0.6981)
(3,-0.6981)
(4,-0.6981)
(5,-0.6981)
};
\end{axis}
\end{tikzpicture}

\vspace{0.09cm}
\begin{tikzpicture}
\begin{axis}[
name = axis1,
title = {Offensive asymmetric match: \\ Larne vs Gent},
title style = {font=\small,align=center},
xlabel = Value of $m$ (goal difference),
x label style = {font=\small},
x tick label style={/pgf/number format/1000 sep=},
xtick = data,
ylabel = Change in probability,
y label style = {font=\small},
width = 0.5\textwidth,
height = 0.38\textwidth,
ymajorgrids = true,
xmin = -5.2,
xmax = 5.2,
ymin = -1.05,
ymax = 1.05,
] 
\addplot [red, thick, dashed, mark=o, mark size=2pt, mark options=solid] coordinates {
(-5,0)
(-4,0)
(-3,0)
(-2,0)
(-1,0)
(0,0)
(1,0)
(2,0)
(3,0)
(4,0)
(5,0)
};
\addplot [red, very thick, dotted, mark=none] coordinates {
(-5,0)
(-4,0)
(-3,0)
(-2,0)
(-1,0)
(0,0)
(1,0)
(2,0)
(3,0)
(4,0)
(5,0)
};
\addplot [blue, thick, dashdotted, mark=asterisk, mark size=2.5pt, mark options={solid,semithick}] coordinates {
(-5,0.6152)
(-4,0.5845)
(-3,0.5291)
(-2,0.4585)
(-1,0.3254)
(0,0)
(1,0)
(2,0)
(3,0)
(4,0)
(5,0)
};
\addplot [blue, very thick, dotted, mark=none] coordinates {
(-5,0.4615)
(-4,0.4615)
(-3,0.4615)
(-2,0.4615)
(-1,0.4615)
(0,0)
(1,0)
(2,0)
(3,0)
(4,0)
(5,0)
};
\end{axis}

\begin{axis}[
at = {(axis1.south east)},
xshift = 0.08\textwidth,
title = {Defensive asymmetric match: \\ LASK vs V\'ikingur Reykjav\'ik},
title style = {font=\small,align=center},
xlabel = Value of $m$ (goal difference),
x label style = {font=\small},
x tick label style={/pgf/number format/1000 sep=},
xtick = data,
width = 0.5\textwidth,
height = 0.38\textwidth,
ymajorgrids = true,
xmin = -5.2,
xmax = 5.2,
ymin = -1.05,
ymax = 1.05,
] 
\addplot [red, thick, dashed, mark=o, mark size=2pt, mark options=solid] coordinates {
(-5,0)
(-4,0)
(-3,0)
(-2,0)
(-1,0)
(0,0)
(1,0)
(2,0)
(3,0)
(4,0)
(5,0)
};
\addplot [red, very thick, dotted, mark=none] coordinates {
(-5,0)
(-4,0)
(-3,0)
(-2,0)
(-1,0)
(0,0)
(1,0)
(2,0)
(3,0)
(4,0)
(5,0)
};
\addplot [blue, thick, dashdotted, mark=asterisk, mark size=2.5pt, mark options={solid,semithick}] coordinates {
(-5,0)
(-4,0)
(-3,0)
(-2,0)
(-1,0)
(0,0)
(1,-0.0842000000000001)
(2,-0.1816)
(3,-0.2761)
(4,-0.3752)
(5,-0.4482)
};
\addplot [blue, very thick, dotted, mark=none] coordinates {
(-5,0)
(-4,0)
(-3,0)
(-2,0)
(-1,0)
(0,0)
(1,-0.2308)
(2,-0.2308)
(3,-0.2308)
(4,-0.2308)
(5,-0.2308)
};
\end{axis}

\end{tikzpicture}

\vspace{0.09cm}
\begin{tikzpicture}
\begin{axis}[
name = axis1,
title = {Antagonistic match: \\ Molde vs Mlad\'a Boleslav},
title style = {font=\small,align=center},
xlabel = Value of $m$ (goal difference),
x label style = {font=\small},
x tick label style={/pgf/number format/1000 sep=},
xtick = data,
ylabel = Change in probability,
y label style = {font=\small},
width = 0.5\textwidth,
height = 0.38\textwidth,
ymajorgrids = true,
xmin = -5.2,
xmax = 5.2,
ymin = -1.05,
ymax = 1.05,
] 
\addplot [red, thick, dashed, mark=o, mark size=2pt, mark options=solid] coordinates {
(-5,0)
(-4,0)
(-3,0)
(-2,0)
(-1,0)
(0,0)
(1,0.9841)
(2,0.9978)
(3,0.9999)
(4,0.9999)
(5,1)
};
\addplot [red, very thick, dotted, mark=none] coordinates {
(-5,0)
(-4,0)
(-3,0)
(-2,0)
(-1,0)
(0,0)
(1,0.9933)
(2,0.9933)
(3,0.9933)
(4,0.9933)
(5,0.9933)
};
\addplot [blue, thick, dashdotted, mark=asterisk, mark size=2.5pt, mark options={solid,semithick}] coordinates {
(-5,0.0633)
(-4,0.0633)
(-3,0.0633)
(-2,0.0633)
(-1,0.0633)
(0,0)
(1,-0.9076)
(2,-0.9201)
(3,-0.9235)
(4,-0.9246)
(5,-0.9246)
};
\addplot [blue, very thick, dotted, mark=none] coordinates {
(-5,0.0633)
(-4,0.0633)
(-3,0.0633)
(-2,0.0633)
(-1,0.0633)
(0,0)
(1,-0.9172)
(2,-0.9172)
(3,-0.9172)
(4,-0.9172)
(5,-0.9172)
};
\end{axis}

\begin{axis}[
at = {(axis1.south east)},
xshift = 0.08\textwidth,
title = {Antagonistic match: \\ Real Betis vs HJK},
title style = {font=\small,align=center},
xlabel = Value of $m$ (goal difference),
x label style = {font=\small},
x tick label style={/pgf/number format/1000 sep=},
xtick = data,
width = 0.5\textwidth,
height = 0.38\textwidth,
ymajorgrids = true,
xmin = -5.2,
xmax = 5.2,
ymin = -1.05,
ymax = 1.05,
legend style = {font=\small,at={(-0.65,-0.25)},anchor=north west,legend columns=2},
legend entries = {Home team$\qquad$,Away team}
] 
\addplot [red, thick, dashed, mark=o, mark size=2pt, mark options=solid] coordinates {
(-5,-0.6972)
(-4,-0.586)
(-3,-0.456)
(-2,-0.0516)
(-1,-0.0148)
(0,0)
(1,0)
(2,0)
(3,0)
(4,0)
(5,0)
};
\addplot [blue, thick, dashdotted, mark=asterisk, mark size=2.5pt, mark options={solid,semithick}] coordinates {
(-5,0.9991)
(-4,0.9891)
(-3,0.9633)
(-2,0.5475)
(-1,0.359)
(0,0)
(1,0)
(2,0)
(3,0)
(4,0)
(5,0)
};
\addplot [red, very thick, dotted, mark=none] coordinates {
(-5,-0.0401)
(-4,-0.0401)
(-3,-0.0401)
(-2,-0.0401)
(-1,-0.0401)
(0,0)
(1,0)
(2,0)
(3,0)
(4,0)
(5,0)
};
\addplot [blue, very thick, dotted, mark=none] coordinates {
(-5,0.4213)
(-4,0.4213)
(-3,0.4213)
(-2,0.4213)
(-1,0.4213)
(0,0)
(1,0)
(2,0)
(3,0)
(4,0)
(5,0)
};
\end{axis}
\end{tikzpicture}

\captionsetup{justification=centering}
\caption{Incentives in six representative matches played in the \\ last round of the 2024/25 UEFA Conference League league phase \\ \vspace{0.2cm}
\footnotesize{\emph{Note}: The dotted line shows the (average) change when the value of $m$ is simulated.}}
\label{Fig1}
\end{figure}


Figure~\ref{Fig1} plots the possible gain $\mathcal{G}_i$ and loss $\mathcal{L}_i$ for six matches (12 teams) as a function of parameter $m$ based on 10 thousand simulation runs. $\mathcal{G}_i$ and $\mathcal{L}_i$ are also depicted for the expected value of $m$, shown by the dotted lines, which overlap with the values for a given $m$ if the latter is independent of $m$.
For example, if APOEL wins by 1-0 against Astana, the probability that APOEL (Astana) qualifies directly for the Round of 16 (the knockout phase play-offs) increases by 95.75\% (does not change).
We give a detailed analysis of three matches based on Tables~\ref{Table1}--\ref{Table3}.

First, APOEL was ranked 10th before the last round, and the most favourable position that can be achieved by playing a draw of 0-0 is the 8th: the first five teams, as well as one team playing in the matches Jagiellonia Bia{\l}ystok  vs Olimpija Ljubljana and Rapid Wien vs Copenhagen, respectively, are guaranteed to be ranked higher than APOEL if it plays 0-0. Therefore, APOEL can be the 8th only if Shamrock Rovers loses against Chelsea with a goal margin of at least 5.
On the other hand, it is quite likely that at least two higher ranked teams are overtaken if APOEL wins. Last but not least, APOEL is guaranteed to finish in the top 24 even with a loss.
The situation of the 28th Astana is even clearer. A draw of 0-0 is equivalent to a loss as the first 23 teams and one team from the match TSC vs Noah will be ahead of Astana. However, even a small victory of 1-0 makes elimination less likely than finishing in the top 24.
Consequently, both clubs lose almost nothing by conceding some goals, while they have strong incentives to win. This leads to a game where no team should play defensively; the match is called offensive, and the strength of incentives $\kappa$ is high, exceeding 65.

Second, the 20th Borac Banja Luka and the 21st Omonia play against each other. None of them could finish in the top 8: the first six teams and one team playing in the matches Jagiellonia Bia{\l}ystok vs Olimpija Ljubljana and Rapid Wien vs Copenhagen, respectively, are guaranteed to be ranked higher than the winner of this match. However, the loser is certainly overtaken by the winner, as well as one team playing in the match Molde vs Mlad\'a Boleslav. Thus, Borac Banja Luka is eliminated if at least two teams from a set of four (Heart of Midlothian, {\. I}stanbul Ba{\c s}ak{\c s}ehir, Celje, TSC) win, while this ``threatening'' set contains four additional teams (Astana, HJK, St.~Gallen, Noah) for Omonia.
Even though, in contrast to Borac Banja Luka, a draw does not guarantee finishing in the top 24 for Omonia, its chance remains remarkably high due to the negative goal difference of all lower-ranked teams.
Therefore, a win of 1-0 does not change (increases) the probability of qualification (by only 2.03\%) for Borac Banja Luka (Omonia) compared to playing a goalless draw, while a loss of 0-1 reduces it by more than 50 percentage points for both teams.
To conclude, they win almost nothing by scoring some goals, but have robust incentives to avoid a loss, with the value of $\kappa$ approaching 55. This probably leads to a boring game, where no team should play offensively; a similar match is called defensive. Note also that the top right plot (Borac Banja Luka vs Omonia) of Figure~\ref{Fig1} is essentially a reflection of the top left plot (APOEL vs Astana) across the point (0,0).

Third, the 22nd Mlad\'a Boleslav plays against the 26th Molde. Analogous to Omonia, a draw of 0-0 is quite favourable for Mlad\'a Boleslav despite its worse goal difference, while a win barely increases its probability of finishing in the top 24. On the other hand, a draw of 0-0 is worthless for Molde because the first 23 teams and one team playing in the match Celje vs The New Saints will certainly be ranked higher. But even a minimal win of 1-0 means qualification with a probability of 98.41\%, and this outcome increases the chance of Mlad\'a Boleslav being eliminated by more than 90 percentage points.
Consequently, Mlad\'a Boleslav wins essentially nothing by scoring some goals, but gains much by avoiding a loss. In contrast, Molde does not lose anything by conceding some goals, and has powerful incentives to win. The spectators should prepare for a match where one team uses an offensive, and the other uses a defensive strategy; the match is strongly antagonistic with $\kappa$ above 85.

The match Real Betis vs HJK is perhaps the most difficult to categorise. HJK should win; otherwise, the first 24 teams will be ranked ahead of it. Betis has the best goal difference among all teams that collected 7 points; even a small loss does not threaten its chances of qualification. Therefore, the match would be offensive asymmetric if $m=1$ is assumed, but Betis should avoid a larger defeat, so the match becomes antagonistic with the expected value of $m$. Nevertheless, the incentives are clearly weaker than in the previous match Molde vs Mlad\'a Boleslav, reflected by the smaller value of $\kappa$, which remains around 4.

Finally, Figure~\ref{Fig1} presents two matches where one team is already eliminated. In the case of Larne vs Gent, the home team cannot reach the top 24, while the away team finishes among the best eight with a decent chance if it wins. Hence, this is an offensive asymmetric match: scoring a goal is much more important than avoiding a loss for the team that is not indifferent, indicated by the high value of $\kappa$, which is above 45.
The counterpart is the match LASK vs V\'ikingur Reykjav\'ik, where the away team has nothing to gain by winning but could be easily eliminated if it loses. Thus, it is called a defensive asymmetric match. Again, the corresponding middle right plot is a reflection of the middle left plot across the point (0,0).

The characteristics of the remaining 12 matches are listed in Table~\ref{Table2} and plotted in Figures~\ref{Fig_A1} and \ref{Fig_A2} in the Appendix.
Three further matches are defensive asymmetric, but there is no unimportant match. Six additional matches are offensive; the value of $\kappa$ exceeds 25 for three of them. No other defensive match exists. One of the three remaining antagonistic matches, Lugano vs Pafos, has a $\kappa$ above 65, showing an extremely strong clash between the incentives of the opposing teams. 




\subsection{The distribution of match types in UEFA club competitions} \label{Sec42}

\input{Figure2_UEFA_CL_2023_groups_Elo}

Obviously, the proportion of matches in the six categories highly depends on competitive balance: if the strengths of the teams are closer, more draws and intransitive cycles are expected, which increases the chance that the teams are closer before the last round. Our simulations of the earlier group stage format provide important insights into this issue, as the eight groups in the 2023/24 UEFA Champions League are surprisingly heterogeneous, see Figure~\ref{Fig2}. For instance, the range of Elo ratings is below 100 in Group F, but approaches 500 in Group G such that the difference between both the two strongest and the two middle teams is around 200-250. The uniqueness of Group B resides in the closeness of the three weakest teams, while Groups A and C are similar to each other, except for the difference between the top two teams. Last but not least, Group E contains a clear underdog.

\begin{figure}[t!]
\centering

\begin{tikzpicture}
\begin{axis}[
name = axis1,
title = {Unimportant match},
title style = {font=\small},
width = 0.48\textwidth, 
height = 0.43\textwidth,
xmajorgrids,
ymajorgrids,
scaled x ticks = false,
xlabel = {Probability (\%)},
xlabel style = {align=center, font=\small},
xticklabel style = {/pgf/number format/fixed,/pgf/number format/precision=5},
xmin = 0,
xbar stacked,
bar width = 6pt,
symbolic y coords = {Group A,Group B,Group C,Group D,Group E,Group F,Group G,Group H,CL 2024,CL 2025,EL 2024,EL 2025,KL 2024,KL 2025},
ytick = data,
y dir = reverse,
enlarge y limits = 0.05,
]
\addplot [blue, thick, pattern = crosshatch dots, pattern color = blue] coordinates{
(25.6916666666667,Group A)
(17.125,Group B)
(21.35,Group C)
(16.1916666666667,Group D)
(19.6916666666667,Group E)
(14.2333333333333,Group F)
(27.9583333333333,Group G)
(24.925,Group H)
(10.9111111111111,CL 2024)
(10.9111111111111,CL 2025)
(10.0111111111111,EL 2024)
(9.54444444444444,EL 2025)
(6.72222222222222,KL 2024)
(6.13333333333333,KL 2025)
};
\end{axis}

\begin{axis}[
at = {(axis1.south east)},
xshift = 0.115\textwidth,
title = {Defensive asymmetric match},
title style = {font=\small},
width = 0.48\textwidth, 
height = 0.43\textwidth,
xmajorgrids,
ymajorgrids,
scaled x ticks = false,
xlabel = {Probability (\%)},
xlabel style = {align=center, font=\small},
xticklabel style = {/pgf/number format/fixed,/pgf/number format/precision=5},
xmin = 0,
xbar stacked,
bar width = 6pt,
symbolic y coords = {Group A,Group B,Group C,Group D,Group E,Group F,Group G,Group H,CL 2024,CL 2025,EL 2024,EL 2025,KL 2024,KL 2025},
ytick = data,
y dir = reverse,
enlarge y limits = 0.05,
]
\addplot [blue, thick, pattern = crosshatch dots, pattern color = blue] coordinates{
(7.65833333333333,Group A)
(6.85833333333333,Group B)
(7.31666666666667,Group C)
(7.03333333333333,Group D)
(6.54166666666667,Group E)
(6.38333333333333,Group F)
(8.24166666666667,Group G)
(7.16666666666667,Group H)
(9.85555555555556,CL 2024)
(9.75555555555556,CL 2025)
(8.55555555555556,EL 2024)
(9.24444444444444,EL 2025)
(7.83333333333333,KL 2024)
(8.28888888888889,KL 2025)
};
\addplot [red, thick, pattern = vertical lines, pattern color = red] coordinates{
(8.525,Group A)
(7.125,Group B)
(7.6,Group C)
(6.81666666666667,Group D)
(6.85833333333333,Group E)
(6.09166666666667,Group F)
(9.53333333333333,Group G)
(8.56666666666667,Group H)
(6.44444444444444,CL 2024)
(5.98888888888889,CL 2025)
(6.36666666666667,EL 2024)
(6.65555555555556,EL 2025)
(4.72222222222222,KL 2024)
(4.73333333333333,KL 2025)
};
\end{axis}
\end{tikzpicture}

\vspace{0.09cm}
\begin{tikzpicture}
\begin{axis}[
name = axis1,
title = {Offensive asymmetric match},
title style = {font=\small},
width = 0.48\textwidth, 
height = 0.43\textwidth,
xmajorgrids,
ymajorgrids,
scaled x ticks = false,
xlabel = {Probability (\%)},
xlabel style = {align=center, font=\small},
xticklabel style = {/pgf/number format/fixed,/pgf/number format/precision=5},
xmin = 0,
xbar stacked,
bar width = 6pt,
symbolic y coords = {Group A,Group B,Group C,Group D,Group E,Group F,Group G,Group H,CL 2024,CL 2025,EL 2024,EL 2025,KL 2024,KL 2025},
ytick = data,
y dir = reverse,
enlarge y limits = 0.05,
]
\addplot [blue, thick, pattern = crosshatch dots, pattern color = blue] coordinates{
(14.3333333333333,Group A)
(16.1166666666667,Group B)
(15.4166666666667,Group C)
(16.75,Group D)
(16.3583333333333,Group E)
(16,Group F)
(14.025,Group G)
(13.95,Group H)
(18.9,CL 2024)
(18.2222222222222,CL 2025)
(18.1666666666667,EL 2024)
(18.3777777777778,EL 2025)
(16.9111111111111,KL 2024)
(16.9555555555556,KL 2025)
};
\addplot [red, thick, pattern = vertical lines, pattern color = red] coordinates{
(8.8,Group A)
(9.80833333333333,Group B)
(9.70833333333333,Group C)
(9.25833333333333,Group D)
(9.925,Group E)
(9.66666666666667,Group F)
(9.15833333333333,Group G)
(9.35,Group H)
(7.01111111111111,CL 2024)
(6.44444444444444,CL 2025)
(6.66666666666667,EL 2024)
(6.96666666666667,EL 2025)
(5.87777777777778,KL 2024)
(5.36666666666667,KL 2025)
};
\end{axis}

\begin{axis}[
at = {(axis1.south east)},
xshift = 0.115\textwidth,
title = {Antagonistic match},
title style = {font=\small},
width = 0.48\textwidth, 
height = 0.43\textwidth,
xmajorgrids,
ymajorgrids,
scaled x ticks = false,
xlabel = {Probability (\%)},
xlabel style = {align=center, font=\small},
xticklabel style = {/pgf/number format/fixed,/pgf/number format/precision=5},
xmin = 0,
xbar stacked,
bar width = 6pt,
symbolic y coords = {Group A,Group B,Group C,Group D,Group E,Group F,Group G,Group H,CL 2024,CL 2025,EL 2024,EL 2025,KL 2024,KL 2025},
ytick = data,
y dir = reverse,
enlarge y limits = 0.05,
]
\addplot [blue, thick, pattern = crosshatch dots, pattern color = blue] coordinates{
(26.7916666666667,Group A)
(31.5083333333333,Group B)
(29.75,Group C)
(31.8333333333333,Group D)
(30.2333333333333,Group E)
(32.6916666666667,Group F)
(24.2416666666667,Group G)
(27.55,Group H)
(11.3111111111111,CL 2024)
(12.3333333333333,CL 2025)
(12.6555555555556,EL 2024)
(12.2444444444444,EL 2025)
(15.0666666666667,KL 2024)
(15.1777777777778,KL 2025)
};
\addplot [red, thick, pattern = vertical lines, pattern color = red] coordinates{
(5.35833333333333,Group A)
(6.59166666666667,Group B)
(5.21666666666667,Group C)
(7.03333333333333,Group D)
(6.01666666666667,Group E)
(8.8,Group F)
(5,Group G)
(5.30833333333333,Group H)
(11.3777777777778,CL 2024)
(12.0777777777778,CL 2025)
(12.5555555555556,EL 2024)
(11.5777777777778,EL 2025)
(12.7888888888889,KL 2024)
(12.3888888888889,KL 2025)
};
\end{axis}
\end{tikzpicture}

\vspace{0.09cm}
\begin{tikzpicture}
\begin{axis}[
name = axis1,
title = {Defensive match},
title style = {font=\small},
width = 0.48\textwidth, 
height = 0.43\textwidth,
xmajorgrids,
ymajorgrids,
scaled x ticks = false,
xlabel = {Probability (\%)},
xlabel style = {align=center, font=\small},
xticklabel style = {/pgf/number format/fixed,/pgf/number format/precision=5},
xmin = 0,
xbar stacked,
bar width = 6pt,
symbolic y coords = {Group A,Group B,Group C,Group D,Group E,Group F,Group G,Group H,CL 2024,CL 2025,EL 2024,EL 2025,KL 2024,KL 2025},
ytick = data,
y dir = reverse,
enlarge y limits = 0.05,
]
\addplot [blue, thick, pattern = crosshatch dots, pattern color = blue] coordinates{
(0.366666666666667,Group A)
(0.491666666666667,Group B)
(0.416666666666667,Group C)
(0.533333333333333,Group D)
(0.466666666666667,Group E)
(0.483333333333333,Group F)
(0.2,Group G)
(0.375,Group H)
(2.72222222222222,CL 2024)
(2.05555555555556,CL 2025)
(2.66666666666667,EL 2024)
(2.55555555555556,EL 2025)
(3.37777777777778,KL 2024)
(3.26666666666667,KL 2025)
};
\addplot [red, thick, pattern = vertical lines, pattern color = red] coordinates{
(0.666666666666667,Group A)
(1.14166666666667,Group B)
(0.85,Group C)
(1.125,Group D)
(1.025,Group E)
(1.28333333333333,Group F)
(0.508333333333333,Group G)
(0.766666666666667,Group H)
(3.96666666666667,CL 2024)
(4.2,CL 2025)
(4.25555555555556,EL 2024)
(4.53333333333333,EL 2025)
(4.74444444444444,KL 2024)
(4.42222222222222,KL 2025)
};
\end{axis}

\begin{axis}[
at = {(axis1.south east)},
xshift = 0.115\textwidth,
title = {Offensive match},
title style = {font=\small},
width = 0.48\textwidth, 
height = 0.43\textwidth,
xmajorgrids,
ymajorgrids,
scaled x ticks = false,
xlabel = {Probability (\%)},
xlabel style = {align=center, font=\small},
xticklabel style = {/pgf/number format/fixed,/pgf/number format/precision=5},
xmin = 0,
xbar stacked,
bar width = 6pt,
symbolic y coords = {Group A,Group B,Group C,Group D,Group E,Group F,Group G,Group H,CL 2024,CL 2025,EL 2024,EL 2025,KL 2024,KL 2025},
ytick = data,
y dir = reverse,
enlarge y limits = 0.05,
legend style = {font=\small,at={(-0.9,-0.25)},anchor=north west,legend columns=2},
legend entries = {Strong incentives$\qquad$, Moderate incentives},
]
\addplot [blue, thick, pattern = crosshatch dots, pattern color = blue] coordinates{
(0.233333333333333,Group A)
(0.566666666666667,Group B)
(0.391666666666667,Group C)
(0.475,Group D)
(0.466666666666667,Group E)
(0.525,Group F)
(0.158333333333333,Group G)
(0.275,Group H)
(9.62222222222222,CL 2024)
(9.47777777777778,CL 2025)
(9.94444444444444,EL 2024)
(9.77777777777778,EL 2025)
(13.1333333333333,KL 2024)
(13.2888888888889,KL 2025)
};
\addplot [red, thick, pattern = vertical lines, pattern color = red] coordinates{
(1.575,Group A)
(2.66666666666667,Group B)
(1.98333333333333,Group C)
(2.95,Group D)
(2.41666666666667,Group E)
(3.84166666666667,Group F)
(0.975,Group G)
(1.76666666666667,Group H)
(7.87777777777778,CL 2024)
(8.53333333333333,CL 2025)
(8.15555555555556,EL 2024)
(8.52222222222222,EL 2025)
(8.82222222222222,KL 2024)
(9.97777777777778,KL 2025)
};
\end{axis}
\end{tikzpicture}

\captionsetup{justification=centerfirst}
\caption{The proportion of different categories of matches in UEFA club competitions}
\label{Fig3}

\end{figure}


Figure~\ref{Fig3} shows the relative frequency of the six types of matches for these eight groups, as well as for the six seasons of UEFA club competitions played in an incomplete round-robin format. The proportion of unimportant matches is in close association with group balance; this is halved to 17\% in the most balanced Group F compared to the highly unbalanced Group G. Nonetheless, unimportant matches occur even more rarely in the novel league phase, as their probability always remains below 11\%.

Regarding asymmetric matches, the differences among the 14 scenarios are smaller than for unimportant matches. Nonetheless, defensive asymmetric matches are more likely in unbalanced groups, and the Conference League seems to contain fewer games with one indifferent team.

Antagonistic matches are more frequent in the previous group stage. Their proportion is higher if a group is more balanced, or if the incomplete round-robin tournament consists of fewer rounds. The latter relationship also holds for antagonistic matches with strong incentives.
Although antagonistic matches are competitive, some spectators might not be enthusiastic to see a game where one team plays offensively and the other plays defensively.

The effects of the incomplete round-robin format are most visible for defensive and offensive matches. The relative frequency of a defensive match is increased from 2\% in a balanced group to at least 6\%, and sometimes to 8\%. This threatens the integrity of UEFA club competitions by inspiring collusive behaviour---recall that the match Borac Banja Luka vs Omonia, studied in Section~\ref{Sec41}, indeed resulted in a goalless draw. Furthermore, while defensive matches with strong incentives to play a draw ($\kappa$ above 25) are essentially non-existent with a probability of at most 0.5\% in the group stage, now their proportion is 2\% even in the Champions League.

On the other hand, the new format of UEFA club competitions also increases the proportion of matches in the last round where both teams should play offensively, from less than 4.5\% to more than 17\%. In addition, while the previous group stage contains strongly offensive matches only occasionally, about a tenth of all matches are offensive with strong incentives to win ($\kappa$ at least 25) in the Champions League, and even higher in the Conference League.

\input{Figure4_match_distribution_competitive}

Finally, Figure~\ref{Fig4} aims to check the official statement of UEFA---cited at the beginning of the paper---that the new format leads to more competitive matches. Among the six categories, antagonistic and offensive matches can be called competitive, as in asymmetric and unimportant matches, at least one team is indifferent, while defensive matches could inspire collusion. According to our results, even though an incomplete round-robin league with six rounds clearly outperforms even a balanced group from this perspective, the advantage of the incomplete round-robin league with eight rounds remains questionable.

However, Figure~\ref{Fig4} does not reflect the substantial change in the composition of competitive matches towards offensive games at the expense of antagonistic games. Thus, if UEFA aimed to increase the proportion of matches where both teams should exert more effort to attack than to defend, then the reform is certainly successful. Nevertheless, this is achieved at the cost of more defensive matches that can lead to serious scandals.

\subsection{Sensitivity analysis} \label{Sec43}

\begin{figure}[t!]
\centering

\begin{tikzpicture}
\begin{axis}[
name = axis1,
title = {Unimportant match},
title style = {font=\small},
width = 0.48\textwidth, 
height = 0.25\textwidth,
xmajorgrids,
ymajorgrids,
scaled x ticks = false,
xlabel = {Probability (\%)},
xlabel style = {align=center, font=\small},
xticklabel style = {/pgf/number format/fixed,/pgf/number format/precision=5},
xmin = 0,
bar width = 6pt,
symbolic y coords = {Group F,Group G,CL 2024,KL 2024},
ytick = data,
y dir = reverse,
enlarge y limits = 0.1,
]
\addplot [ForestGreen, only marks, mark = triangle, very thick] coordinates{
(13.9916666666667,Group F)
(27.3583333333333,Group G)
(5.76666666666667,CL 2024)
(3.02222222222222,KL 2024)
};
\addplot [red, only marks, mark = star, thick] coordinates{
(14.0916666666667,Group F)
(27.6166666666667,Group G)
(8.13333333333333,CL 2024)
(4.6,KL 2024)
};
\addplot [blue, only marks, mark = square, very thick] coordinates{
(14.2333333333333,Group F)
(27.9583333333333,Group G)
(10.9111111111111,CL 2024)
(6.72222222222222,KL 2024)
};
\addplot [brown, only marks, mark = pentagon, very thick] coordinates{
(14.475,Group F)
(29.8166666666667,Group G)
(14.2666666666667,CL 2024)
(9.13333333333333,KL 2024)
};
\end{axis}

\begin{axis}[
at = {(axis1.south east)},
xshift = 0.12\textwidth,
title = {Defensive asymmetric match},
title style = {font=\small},
width = 0.48\textwidth, 
height = 0.25\textwidth,
xmajorgrids,
ymajorgrids,
scaled x ticks = false,
xlabel = {Probability (\%)},
xlabel style = {align=center, font=\small},
xticklabel style = {/pgf/number format/fixed,/pgf/number format/precision=5},
xmin = 0,
bar width = 6pt,
symbolic y coords = {Group F,Group G,CL 2024,KL 2024},
ytick = data,
y dir = reverse,
enlarge y limits = 0.1,
]
\addplot [ForestGreen, only marks, mark = triangle, very thick] coordinates{
(13.0166666666667,Group F)
(18.425,Group G)
(13.7111111111111,CL 2024)
(10.0444444444444,KL 2024)
};
\addplot [red, only marks, mark = star, thick] coordinates{
(12.775,Group F)
(18.275,Group G)
(15.1444444444444,CL 2024)
(11.4,KL 2024)
};
\addplot [blue, only marks, mark = square, very thick] coordinates{
(12.475,Group F)
(17.775,Group G)
(16.3,CL 2024)
(12.5555555555556,KL 2024)
};
\addplot [brown, only marks, mark = pentagon, very thick] coordinates{
(11.1666666666667,Group F)
(15.9916666666667,Group G)
(16.4111111111111,CL 2024)
(13.3777777777778,KL 2024)
};
\end{axis}
\end{tikzpicture}

\vspace{0.2cm}
\begin{tikzpicture}
\begin{axis}[
name = axis1,
title = {Offensive asymmetric match},
title style = {font=\small},
width = 0.48\textwidth, 
height = 0.25\textwidth,
xmajorgrids,
ymajorgrids,
scaled x ticks = false,
xlabel = {Probability (\%)},
xlabel style = {align=center, font=\small},
xticklabel style = {/pgf/number format/fixed,/pgf/number format/precision=5},
xmin = 0,
bar width = 6pt,
symbolic y coords = {Group F,Group G,CL 2024,KL 2024},
ytick = data,
y dir = reverse,
enlarge y limits = 0.1,
]
\addplot [ForestGreen, only marks, mark = triangle, very thick] coordinates{
(24.9333333333333,Group F)
(22.575,Group G)
(20.9333333333333,CL 2024)
(17.2777777777778,KL 2024)
};
\addplot [red, only marks, mark = star, thick] coordinates{
(25.1833333333333,Group F)
(22.7083333333333,Group G)
(23.6666666666667,CL 2024)
(19.8222222222222,KL 2024)
};
\addplot [blue, only marks, mark = square, very thick] coordinates{
(25.6666666666667,Group F)
(23.1833333333333,Group G)
(25.9111111111111,CL 2024)
(22.7888888888889,KL 2024)
};
\addplot [brown, only marks, mark = pentagon, very thick] coordinates{
(27.1166666666667,Group F)
(23.775,Group G)
(27.9111111111111,CL 2024)
(25.3,KL 2024)
};
\end{axis}

\begin{axis}[
at = {(axis1.south east)},
xshift = 0.12\textwidth,
title = {Antagonistic match},
title style = {font=\small},
width = 0.48\textwidth, 
height = 0.25\textwidth,
xmajorgrids,
ymajorgrids,
scaled x ticks = false,
xlabel = {Probability (\%)},
xlabel style = {align=center, font=\small},
xticklabel style = {/pgf/number format/fixed,/pgf/number format/precision=5},
xmin = 0,
bar width = 6pt,
symbolic y coords = {Group F,Group G,CL 2024,KL 2024},
ytick = data,
y dir = reverse,
enlarge y limits = 0.1,
]
\addplot [ForestGreen, only marks, mark = triangle, very thick] coordinates{
(41.8416666666667,Group F)
(29.6083333333333,Group G)
(28.8555555555556,CL 2024)
(33.7222222222222,KL 2024)
};
\addplot [red, only marks, mark = star, thick] coordinates{
(41.775,Group F)
(29.45,Group G)
(25.7555555555556,CL 2024)
(31.2111111111111,KL 2024)
};
\addplot [blue, only marks, mark = square, very thick] coordinates{
(41.4916666666667,Group F)
(29.2416666666667,Group G)
(22.6888888888889,CL 2024)
(27.8555555555556,KL 2024)
};
\addplot [brown, only marks, mark = pentagon, very thick] coordinates{
(41.3083333333333,Group F)
(28.5,Group G)
(19.8444444444444,CL 2024)
(24.8888888888889,KL 2024)
};
\end{axis}
\end{tikzpicture}

\vspace{0.2cm}
\begin{tikzpicture}
\begin{axis}[
name = axis1,
title = {Defensive match},
title style = {font=\small},
width = 0.48\textwidth, 
height = 0.25\textwidth,
xmajorgrids,
ymajorgrids,
scaled x ticks = false,
xlabel = {Probability (\%)},
xlabel style = {align=center, font=\small},
xticklabel style = {/pgf/number format/fixed,/pgf/number format/precision=5},
xmin = 0,
bar width = 6pt,
symbolic y coords = {Group F,Group G,CL 2024,KL 2024},
ytick = data,
y dir = reverse,
enlarge y limits = 0.1,
]
\addplot [ForestGreen, only marks, mark = triangle, very thick] coordinates{
(2.11666666666667,Group F)
(0.916666666666667,Group G)
(9.36666666666667,CL 2024)
(10.3,KL 2024)
};
\addplot [red, only marks, mark = star, thick] coordinates{
(2.025,Group F)
(0.85,Group G)
(8.07777777777778,CL 2024)
(9.17777777777778,KL 2024)
};
\addplot [blue, only marks, mark = square, very thick] coordinates{
(1.76666666666667,Group F)
(0.708333333333333,Group G)
(6.68888888888889,CL 2024)
(8.12222222222222,KL 2024)
};
\addplot [brown, only marks, mark = pentagon, very thick] coordinates{
(1.43333333333333,Group F)
(0.616666666666667,Group G)
(5.7,CL 2024)
(7,KL 2024)
};
\end{axis}

\begin{axis}[
at = {(axis1.south east)},
xshift = 0.12\textwidth,
title = {Offensive match},
title style = {font=\small},
width = 0.48\textwidth, 
height = 0.25\textwidth,
xmajorgrids,
ymajorgrids,
scaled x ticks = false,
xlabel = {Probability (\%)},
xlabel style = {align=center, font=\small},
xticklabel style = {/pgf/number format/fixed,/pgf/number format/precision=5},
xmin = 0,
bar width = 6pt,
symbolic y coords = {Group F,Group G,CL 2024,KL 2024},
ytick = data,
y dir = reverse,
enlarge y limits = 0.1,
legend style = {font=\small,at={(-1,-0.5)},anchor=north west,legend columns=4},
legend entries = {$\mathcal{I} = 0 \qquad$, $\mathcal{I} = 0.5\% \qquad$, $\mathcal{I} = 2\% \qquad$, $\mathcal{I} = 5\%$},
]
\addplot [ForestGreen, only marks, mark = triangle, very thick] coordinates{
(4.1,Group F)
(1.11666666666667,Group G)
(21.3666666666667,CL 2024)
(25.6333333333333,KL 2024)
};
\addplot [red, only marks, mark = star, thick] coordinates{
(4.15,Group F)
(1.1,Group G)
(19.2222222222222,CL 2024)
(23.7888888888889,KL 2024)
};
\addplot [blue, only marks, mark = square, very thick] coordinates{
(4.36666666666667,Group F)
(1.13333333333333,Group G)
(17.5,CL 2024)
(21.9555555555556,KL 2024)
};
\addplot [brown, only marks, mark = pentagon, very thick] coordinates{
(4.5,Group F)
(1.3,Group G)
(15.8666666666667,CL 2024)
(20.3,KL 2024)
};
\end{axis}
\end{tikzpicture}

\captionsetup{justification=centerfirst}
\caption{Sensitivity analysis: The impact of the indifference threshold $\mathcal{I}$}
\label{Fig5}

\end{figure}


The simulation framework of Section~\ref{Sec33} has a seemingly arbitrary parameter, the indifference threshold $\mathcal{I}$.
Figure~\ref{Fig5} provides crucial information on its impact for Groups F and G, the least and most unbalanced groups in the 2023/24 UEFA Champions League, as well as for two incomplete round-robin leagues, the 2024/25 UEFA Champions League and the 2024/25 UEFA Conference League. The value of $\mathcal{I}$ essentially does not influence the proportion of match types in the round-robin groups; the only change is the marginally higher (lower) proportion of unimportant (defensive asymmetric) matches in an unbalanced group if the indifference threshold grows.

Contrarily, increasing the threshold from 0 to 5\% more than doubles the proportion of unimportant matches in the novel format, together with a moderate growth in asymmetric matches, while the proportion of the other three match categories is reduced by 21--37\%. In each case, the main conclusions still hold: the incomplete round-robin design implies substantially more defensive and offensive matches. In particular, at least a tenth of matches are switched to offensive, while at least four percent of all matches become defensive compared to a round-robin group. The relative position of the Champions League and the Conference League is entirely unaffected by parameter $\mathcal{I}$, that is, playing fewer rounds magnifies the effects of the reform.

\input{Figure6_sensitivity_group_composition}

The results for the pre-2024 design may be driven by the possibly unique composition of the groups in the 2023/24 UEFA Champions League. Therefore, the proportion of the six categories of matches is considered from another perspective in Figure~\ref{Fig6}.
First, we take the eight groups from the 2022/23 and 2023/24 seasons of the UEFA Champions League, respectively, and determine the minimal and maximal proportions of each match type. They are compared to the proportions observed in the 2024/25 Champions League (incomplete round-robin with eight rounds) and the 2024/25 Conference League (incomplete round-robin with six rounds). The ranges for the 2023/24 Champions League, already seen in Figure~\ref{Fig3}, are wider than the ranges for the previous, 2022/23 season (except for offensive symmetric matches), since the variance in group balance was higher in the 2023/24 season.

To summarise, Figure~\ref{Fig6} reinforces our three main findings:
(1) unimportant matches are less frequent in an incomplete round-robin league;
(2) the proportion of defensive matches greatly increases in relative terms but not so much in absolute terms since defensive matches are rare in a round-robin group;
(3) the proportion of offensive matches robustly increases in both absolute and relative terms compared to even a balanced round-robin group, such as Group F in the 2023/24 Champions League.
In contrast, the observed frequency of other categories (antagonistic, defensive and offensive asymmetric) in the novel format could have occurred in the traditional group stage of UEFA club competitions, with the possible exception of antagonistic matches if the number of rounds is eight. Consequently, the main question is whether the favourable changes in the proportion of unimportant and offensive matches are sufficient to compensate for the higher probability of defensive matches that are vulnerable to (at least tacit) collusion.

\section{Conclusions} \label{Sec5}

The current paper argues for a new probabilistic match classification scheme to resolve two major limitations of the literature: the treatment of indifference in a simple binary sense, and the merging of competitive matches with fundamentally different incentives for the teams.
To that end, we propose a simulation framework to quantify the possible gain and loss from the risky strategy of playing offensively, and apply it to the two recent designs of UEFA club competitions, a round-robin group stage and a single incomplete round-robin league.

The comparison of the two tournament formats has yielded crucial lessons. 
While the smaller proportion of unimportant matches and the substantially higher proportion of offensive matches are clearly favourable developments due to the 2024 reform, the increase in the frequency of defensive matches is worrying. These games create robust incentives for collusion, which is straightforward to implement because each match starts at 0-0. According to Table~\ref{Table1}, no one could have been surprised on 19 December 2024 to see that two teams, Borac Banja Luka and Omonia, played a draw of 0-0. Hence, the official UEFA statement, cited at the beginning of the paper, may be an overstatement: the claim of more competitive matches for \emph{every} club across the board is certainly untrue. As the Disgrace of Gij\'on has shown, seemingly collusive matches can really threaten the integrity of sports \citep[Section~3.9.1]{KendallLenten2017}.

To summarise, we provide robust evidence that the novel classification method is relevant to football; however, it is also applicable to other contests with matches between two contestants. It requires (a) a non-negligible chance of a draw, and (b) the possibility of an offensive (defensive) tactic that increases (decreases) the probability of both winning and losing. Besides football, chess and ice hockey can be promising examples.

Our match classification model can be useful for various stakeholders. Bookmakers could adjust the betting odds by taking the incentives of the contestants into account, and bettors might generate profit from mispriced games. Coaches are allowed to select the starting squad with the highest probability of achieving the beneficial outcome. Media companies can broadcast the most attractive matches for their audience, and spectators  may choose which game to watch from the set of games played simultaneously---both tasks being quite challenging when 18 matches are played at the same time as in the new format of UEFA club competitions.

Hopefully, the proposed model will inspire further simulation and empirical studies. Regarding simulations, it would be interesting to analyse the FIFA World Cup, which has seen several reforms in 2026, including the expansion to 48 teams and less restrictive qualification rules. Comparing the deterministic and probabilistic models can provide more insight into match classification. Our approach is worth applying to all tournament rounds, not just the last.
Future empirical papers should explore the relationship between the incentives created by the matches and the match outcomes. For example, we expect indifferent teams to underperform if their opponent has strong incentives to win. Probably, goals occur later, and (goalless) draws are more frequent in defensive matches than in offensive matches. It needs to be investigated whether the incentives of the teams influence pre-game betting odds and the engagement of the audience.

\section*{Acknowledgements}
\addcontentsline{toc}{section}{Acknowledgements}

We are grateful to three anonymous reviewers and \emph{David Winkelmann} for useful remarks. \\
The research was supported by the National Research, Development and Innovation Office under Grants Advanced 152220 and FK 145838, and the J\'anos Bolyai Research Scholarship of the Hungarian Academy of Sciences.

\bibliographystyle{apalike} 
\bibliography{All_references}

\clearpage
\setcounter{figure}{0}
\renewcommand{\thefigure}{A.\arabic{figure}}

\setcounter{table}{0}
\renewcommand{\thetable}{A.\arabic{table}}

\section*{Appendix}
\addcontentsline{toc}{section}{Appendix}

\begin{table}[ht!]
\centering
\caption{2024/25 UEFA Conference League league phase: standing before the last round}
\label{Table_A1}

\begin{threeparttable}
\rowcolors{3}{}{gray!20}
    \begin{tabularx}{\linewidth}{Cl CCC CCC >{\bfseries}C} \toprule \hiderowcolors
    Pos   & Team  & W     & D     & L     & GF    & GA    & GD    & Pts \\ \bottomrule \showrowcolors
    1     & Chelsea  & 5     & 0     & 0     & 21    & 4     & $+17$    & 15 \\
    2     & Vit\'oria de Guimar{\~ a}es  & 4     & 1     & 0     & 12    & 5     & $+7$     & 13 \\
    3     & Fiorentina  & 4     & 0     & 1     & 17    & 6     & $+11$    & 12 \\
    4     & Legia Warsaw  & 4     & 0     & 1     & 12    & 2     & $+10$    & 12 \\
    5     & Lugano  & 4     & 0     & 1     & 9     & 5     & $+4$     & 12 \\
    6     & Shamrock Rovers  & 3     & 2     & 0     & 11    & 4     & $+7$     & 11 \\
    7     & Cercle Brugge & 3     & 1     & 1     & 13    & 6     & $+7$     & 10 \\
    8     & Jagiellonia Bia{\l}ystok  & 3     & 1     & 1     & 10    & 5     & $+5$     & 10 \\ \hline
    9     & Rapid Wien  & 3     & 1     & 1     & 8     & 5     & $+3$     & 10 \\
    10    & APOEL  & 3     & 1     & 1     & 7     & 4     & $+3$     & 10 \\
    11    & Djurg{\r a}rdens IF  & 3     & 1     & 1     & 8     & 6     & $+2$     & 10 \\
    12    & Pafos  & 3     & 0     & 2     & 9     & 5     & $+4$     & 9 \\
    13    & Olimpija Ljubljana  & 3     & 0     & 2     & 7     & 6     & $+1$     & 9 \\
    14    & Gent  & 3     & 0     & 2     & 8     & 7     & $+1$     & 9 \\
    15    & 1.~FC Heidenheim  & 3     & 0     & 2     & 6     & 6     & 0     & 9 \\
    16    & Copenhagen  & 2     & 2     & 1     & 8     & 6     & $+2$     & 8 \\
    17    & Real Betis  & 2     & 1     & 2     & 5     & 5     & 0     & 7 \\
    18    & Panathinaikos  & 2     & 1     & 2     & 6     & 7     & $-1$    & 7 \\
    19    & V\'ikingur Reykjav\'ik  & 2     & 1     & 2     & 6     & 7     & $-1$    & 7 \\
    20    & Borac Banja Luka  & 2     & 1     & 2     & 4     & 7     & $-3$    & 7 \\
    21    & Omonia  & 2     & 0     & 3     & 7     & 7     & 0     & 6 \\
    22    & Mlad\'a Boleslav  & 2     & 0     & 3     & 4     & 6     & $-2$    & 6 \\
    23    & Heart of Midlothian  & 2     & 0     & 3     & 4     & 7     & $-3$    & 6 \\
    24    & {\. I}stanbul Ba{\c s}ak{\c s}ehir  & 1     & 2     & 2     & 8     & 11    & $-3$    & 5 \\ \hline
    25    & Celje  & 1     & 1     & 3     & 10    & 11    & $-1$    & 4 \\
    26    & Molde  & 1     & 1     & 3     & 6     & 8     & $-2$    & 4 \\
    27    & TSC   & 1     & 1     & 3     & 6     & 10    & $-4$    & 4 \\
    28    & Astana & 1     & 1     & 3     & 3     & 7     & $-4$    & 4 \\
    29    & HJK   & 1     & 1     & 3     & 3     & 8     & $-5$    & 4 \\
    30    & St.~Gallen  & 1     & 1     & 3     & 9     & 17    & $-8$    & 4 \\
    31    & Noah  & 1     & 1     & 3     & 3     & 12    & $-9$    & 4 \\
    32    & The New Saints  & 1     & 0     & 4     & 3     & 7     & $-4$    & 3 \\
    33    & Dinamo Minsk  & 1     & 0     & 4     & 4     & 9     & $-5$    & 3 \\
    34    & LASK  & 0     & 2     & 3     & 3     & 13    & $-10$   & 2 \\
    35    & Petrocub H{\^ i}nce{\c s}ti  & 0     & 1     & 4     & 2     & 11    & $-9$    & 1 \\
    36    & Larne  & 0     & 0     & 5     & 2     & 12    & $-10$   & 0 \\ \bottomrule
    \end{tabularx}
    
\begin{tablenotes} \footnotesize
\item
\emph{Notes}:
The top eight teams directly qualify for the Round of 16, and the next 16 teams (ranked  from 9th to 24th) qualify for the knockout phase play-offs.
\item
Pos = Position; W = Won; D = Drawn; L = Lost; GF = Goals for; GA = Goals against; GD = Goal difference; Pts = Points. All teams played five matches. 
\end{tablenotes}
\end{threeparttable}
\end{table}

\begin{figure}[t!]

\begin{tikzpicture}
\begin{axis}[
name = axis1,
title = {1.~FC Heidenheim vs St.~Gallen},
title style = {font=\small},
xlabel = Value of $m$ (goal difference),
x label style = {font=\small},
x tick label style={/pgf/number format/1000 sep=},
xtick = data,
ylabel = Change in probability,
y label style = {font=\small},
width = 0.5\textwidth,
height = 0.42\textwidth,
ymajorgrids = true,
xmin = -5.2,
xmax = 5.2,
ymin = -1.05,
ymax = 1.05,
] 
\addplot [red, thick, dashed, mark=o, mark size=2pt, mark options=solid] coordinates {
(-5,0)
(-4,0)
(-3,0)
(-2,0)
(-1,0)
(0,0)
(1,0.2016)
(2,0.2131)
(3,0.3437)
(4,0.4611)
(5,0.5322)
};
\addplot [red, very thick, dotted, mark=none] coordinates {
(-5,0)
(-4,0)
(-3,0)
(-2,0)
(-1,0)
(0,0)
(1,0.2924)
(2,0.2924)
(3,0.2924)
(4,0.2924)
(5,0.2924)
};
\addplot [blue, thick, dashdotted, mark=asterisk, mark size=2.5pt, mark options={solid,semithick}] coordinates {
(-5,0.7508)
(-4,0.474)
(-3,0.3254)
(-2,0.24)
(-1,0.1963)
(0,0)
(1,0)
(2,0)
(3,0)
(4,0)
(5,0)
};
\addplot [blue, very thick, dotted, mark=none] coordinates {
(-5,0.218)
(-4,0.218)
(-3,0.218)
(-2,0.218)
(-1,0.218)
(0,0)
(1,0)
(2,0)
(3,0)
(4,0)
(5,0)
};
\end{axis}

\begin{axis}[
at = {(axis1.south east)},
xshift = 0.08\textwidth,
title = {Celje vs The New Saints},
title style = {font=\small},
xlabel = Value of $m$ (goal difference),
x label style = {font=\small},
x tick label style={/pgf/number format/1000 sep=},
xtick = data,
width = 0.5\textwidth,
height = 0.42\textwidth,
ymajorgrids = true,
xmin = -5.2,
xmax = 5.2,
ymin = -1.05,
ymax = 1.05,
] 
\addplot [red, thick, dashed, mark=o, mark size=2pt, mark options=solid] coordinates {
(-5,-0.0305)
(-4,-0.0305)
(-3,-0.0305)
(-2,-0.0305)
(-1,-0.0305)
(0,0)
(1,0.9695)
(2,0.9695)
(3,0.9695)
(4,0.9695)
(5,0.9695)
};
\addplot [red, very thick, dotted, mark=none] coordinates {
(-5,-0.0305)
(-4,-0.0305)
(-3,-0.0305)
(-2,-0.0305)
(-1,-0.0305)
(0,0)
(1,0.9695)
(2,0.9695)
(3,0.9695)
(4,0.9695)
(5,0.9695)
};
\addplot [blue, thick, dashdotted, mark=asterisk, mark size=2.5pt, mark options={solid,semithick}] coordinates {
(-5,0.2679)
(-4,0.2679)
(-3,0.2095)
(-2,0.1874)
(-1,0.1076)
(0,0)
(1,0)
(2,0)
(3,0)
(4,0)
(5,0)
};
\addplot [blue, very thick, dotted, mark=none] coordinates {
(-5,0.1365)
(-4,0.1365)
(-3,0.1365)
(-2,0.1365)
(-1,0.1365)
(0,0)
(1,0)
(2,0)
(3,0)
(4,0)
(5,0)
};
\end{axis}
\end{tikzpicture}

\vspace{0.1cm}
\begin{tikzpicture}
\begin{axis}[
name = axis1,
title = {Cercle Brugge vs {\. I}stanbul Ba{\c s}ak{\c s}ehir},
title style = {font=\small},
xlabel = Value of $m$ (goal difference),
x label style = {font=\small},
x tick label style={/pgf/number format/1000 sep=},
xtick = data,
ylabel = Change in probability,
y label style = {font=\small},
width = 0.5\textwidth,
height = 0.42\textwidth,
ymajorgrids = true,
xmin = -5.2,
xmax = 5.2,
ymin = -1.05,
ymax = 1.05,
] 
\addplot [red, thick, dashed, mark=o, mark size=2pt, mark options=solid] coordinates {
(-5,-0.144)
(-4,-0.144)
(-3,-0.144)
(-2,-0.144)
(-1,-0.144)
(0,0)
(1,0.856)
(2,0.856)
(3,0.856)
(4,0.856)
(5,0.856)
};
\addplot [red, very thick, dotted, mark=none] coordinates {
(-5,-0.144)
(-4,-0.144)
(-3,-0.144)
(-2,-0.144)
(-1,-0.144)
(0,0)
(1,0.856)
(2,0.856)
(3,0.856)
(4,0.856)
(5,0.856)
};
\addplot [blue, thick, dashdotted, mark=asterisk, mark size=2.5pt, mark options={solid,semithick}] coordinates {
(-5,0.9222)
(-4,0.9222)
(-3,0.9222)
(-2,0.9222)
(-1,0.9222)
(0,0)
(1,-0.0778)
(2,-0.0778)
(3,-0.0778)
(4,-0.0778)
(5,-0.0778)
};
\addplot [blue, very thick, dotted, mark=none] coordinates {
(-5,0.9222)
(-4,0.9222)
(-3,0.9222)
(-2,0.9222)
(-1,0.9222)
(0,0)
(1,-0.0778)
(2,-0.0778)
(3,-0.0778)
(4,-0.0778)
(5,-0.0778)
};
\end{axis}

\begin{axis}[
at = {(axis1.south east)},
xshift = 0.08\textwidth,
title = {Chelsea vs Shamrock Rovers},
title style = {font=\small},
xlabel = Value of $m$ (goal difference),
x label style = {font=\small},
x tick label style={/pgf/number format/1000 sep=},
xtick = data,
width = 0.5\textwidth,
height = 0.42\textwidth,
ymajorgrids = true,
xmin = -5.2,
xmax = 5.2,
ymin = -1.05,
ymax = 1.05,
] 
\addplot [red, thick, dashed, mark=o, mark size=2pt, mark options=solid] coordinates {
(-5,0)
(-4,0)
(-3,0)
(-2,0)
(-1,0)
(0,0)
(1,0)
(2,0)
(3,0)
(4,0)
(5,0)
};
\addplot [red, very thick, dotted, mark=none] coordinates {
(-5,0)
(-4,0)
(-3,0)
(-2,0)
(-1,0)
(0,0)
(1,0)
(2,0)
(3,0)
(4,0)
(5,0)
};
\addplot [blue, thick, dashdotted, mark=asterisk, mark size=2.5pt, mark options={solid,semithick}] coordinates {
(-5,0.3403)
(-4,0.3403)
(-3,0.3403)
(-2,0.3403)
(-1,0.3403)
(0,0)
(1,-0.5914)
(2,-0.6113)
(3,-0.6271)
(4,-0.6328)
(5,-0.6572)
};
\addplot [blue, very thick, dotted, mark=none] coordinates {
(-5,0.3403)
(-4,0.3403)
(-3,0.3403)
(-2,0.3403)
(-1,0.3403)
(0,0)
(1,-0.5914)
(2,-0.6113)
(3,-0.6271)
(4,-0.6328)
(5,-0.6572)
};
\end{axis}
\end{tikzpicture}

\vspace{0.1cm}
\begin{tikzpicture}
\begin{axis}[
name = axis1,
title = {Djurg{\r a}rdens IF vs Legia Warsaw},
title style = {font=\small},
xlabel = Value of $m$ (goal difference),
x label style = {font=\small},
x tick label style={/pgf/number format/1000 sep=},
xtick = data,
ylabel = Change in probability,
y label style = {font=\small},
width = 0.5\textwidth,
height = 0.42\textwidth,
ymajorgrids = true,
xmin = -5.2,
xmax = 5.2,
ymin = -1.05,
ymax = 1.05,
] 
\addplot [red, thick, dashed, mark=o, mark size=2pt, mark options=solid] coordinates {
(-5,-0.0001)
(-4,-0.0001)
(-3,-0.0001)
(-2,-0.0001)
(-1,-0.0001)
(0,0)
(1,0.9727)
(2,0.992)
(3,0.9991)
(4,0.9998)
(5,0.9999)
};
\addplot [red, very thick, dotted, mark=none] coordinates {
(-5,-0.0001)
(-4,-0.0001)
(-3,-0.0001)
(-2,-0.0001)
(-1,-0.0001)
(0,0)
(1,0.9844)
(2,0.9844)
(3,0.9844)
(4,0.9844)
(5,0.9844)
};
\addplot [blue, thick, dashdotted, mark=asterisk, mark size=2.5pt, mark options={solid,semithick}] coordinates {
(-5,0)
(-4,0)
(-3,0)
(-2,0)
(-1,0)
(0,0)
(1,-0.2022)
(2,-0.2095)
(3,-0.2276)
(4,-0.2709)
(5,-0.2929)
};
\addplot [blue, very thick, dotted, mark=none] coordinates {
(-5,0)
(-4,0)
(-3,0)
(-2,0)
(-1,0)
(0,0)
(1,-0.2121)
(2,-0.2121)
(3,-0.2121)
(4,-0.2121)
(5,-0.2121)
};
\end{axis}

\begin{axis}[
at = {(axis1.south east)},
xshift = 0.08\textwidth,
title = {Heart of Midlothian vs Petrocub H{\^ i}nce{\c s}ti},
title style = {font=\small},
xlabel = Value of $m$ (goal difference),
x label style = {font=\small},
x tick label style={/pgf/number format/1000 sep=},
xtick = data,
width = 0.5\textwidth,
height = 0.42\textwidth,
ymajorgrids = true,
xmin = -5.2,
xmax = 5.2,
ymin = -1.05,
ymax = 1.05,
legend style = {font=\small,at={(-0.65,-0.25)},anchor=north west,legend columns=2},
legend entries = {Home team$\qquad$,Away team}
] 
\addplot [red, thick, dashed, mark=o, mark size=2pt, mark options=solid] coordinates {
(-5,-0.7326)
(-4,-0.7316)
(-3,-0.7287)
(-2,-0.7197)
(-1,-0.7037)
(0,0)
(1,0.205)
(2,0.205)
(3,0.205)
(4,0.205)
(5,0.205)
};
\addplot [blue, thick, dashdotted, mark=asterisk, mark size=2.5pt, mark options={solid,semithick}] coordinates {
(-5,0)
(-4,0)
(-3,0)
(-2,0)
(-1,0)
(0,0)
(1,0)
(2,0)
(3,0)
(4,0)
(5,0)
};
\addplot [red, very thick, dotted, mark=none] coordinates {
(-5,-0.7103)
(-4,-0.7103)
(-3,-0.7103)
(-2,-0.7103)
(-1,-0.7103)
(0,0)
(1,0.205)
(2,0.205)
(3,0.205)
(4,0.205)
(5,0.205)
};
\addplot [blue, very thick, dotted, mark=none] coordinates {
(-5,0)
(-4,0)
(-3,0)
(-2,0)
(-1,0)
(0,0)
(1,0)
(2,0)
(3,0)
(4,0)
(5,0)
};
\end{axis}
\end{tikzpicture}

\captionsetup{justification=centering}
\caption{Incentives in further matches played in the last \\ round of the 2024/25 UEFA Conference League league phase I. \\ \vspace{0.2cm}
\footnotesize{\emph{Note}: The dotted line shows the (average) change when the value of $m$ is simulated.}}
\label{Fig_A1}
\end{figure}

\begin{figure}[t!]

\begin{tikzpicture}
\begin{axis}[
name = axis1,
title = {Jagiellonia Bia{\l}ystok vs Olimpija Ljubljana},
title style = {font=\small},
xlabel = Value of $m$ (goal difference),
x label style = {font=\small},
x tick label style={/pgf/number format/1000 sep=},
xtick = data,
ylabel = Change in probability,
y label style = {font=\small},
width = 0.5\textwidth,
height = 0.42\textwidth,
ymajorgrids = true,
xmin = -5.2,
xmax = 5.2,
ymin = -1.05,
ymax = 1.05,
] 
\addplot [red, thick, dashed, mark=o, mark size=2pt, mark options=solid] coordinates {
(-5,-0.0903)
(-4,-0.0903)
(-3,-0.0903)
(-2,-0.0903)
(-1,-0.0903)
(0,0)
(1,0.9094)
(2,0.9097)
(3,0.9097)
(4,0.9097)
(5,0.9097)
};
\addplot [red, very thick, dotted, mark=none] coordinates {
(-5,-0.0903)
(-4,-0.0903)
(-3,-0.0903)
(-2,-0.0903)
(-1,-0.0903)
(0,0)
(1,0.9096)
(2,0.9096)
(3,0.9096)
(4,0.9096)
(5,0.9096)
};
\addplot [blue, thick, dashdotted, mark=asterisk, mark size=2.5pt, mark options={solid,semithick}] coordinates {
(-5,0.7589)
(-4,0.6981)
(-3,0.6201)
(-2,0.4894)
(-1,0.3003)
(0,0)
(1,0)
(2,0)
(3,0)
(4,0)
(5,0)
};
\addplot [blue, very thick, dotted, mark=none] coordinates {
(-5,0.4066)
(-4,0.4066)
(-3,0.4066)
(-2,0.4066)
(-1,0.4066)
(0,0)
(1,0)
(2,0)
(3,0)
(4,0)
(5,0)
};
\end{axis}

\begin{axis}[
at = {(axis1.south east)},
xshift = 0.08\textwidth,
title = {Lugano vs Pafos},
title style = {font=\small},
xlabel = Value of $m$ (goal difference),
x label style = {font=\small},
x tick label style={/pgf/number format/1000 sep=},
xtick = data,
width = 0.5\textwidth,
height = 0.42\textwidth,
ymajorgrids = true,
xmin = -5.2,
xmax = 5.2,
ymin = -1.05,
ymax = 1.05,
] 
\addplot [red, thick, dashed, mark=o, mark size=2pt, mark options=solid] coordinates {
(-5,-0.8617)
(-4,-0.8617)
(-3,-0.8204)
(-2,-0.7306)
(-1,-0.6514)
(0,0)
(1,0.0233)
(2,0.0233)
(3,0.0233)
(4,0.0233)
(5,0.0233)
};
\addplot [red, very thick, dotted, mark=none] coordinates {
(-5,-0.6954)
(-4,-0.6954)
(-3,-0.6954)
(-2,-0.6954)
(-1,-0.6954)
(0,0)
(1,0.0233)
(2,0.0233)
(3,0.0233)
(4,0.0233)
(5,0.0233)
};
\addplot [blue, thick, dashdotted, mark=asterisk, mark size=2.5pt, mark options={solid,semithick}] coordinates {
(-5,0.952)
(-4,0.9024)
(-3,0.8656)
(-2,0.8222)
(-1,0.7904)
(0,0)
(1,0)
(2,0)
(3,0)
(4,0)
(5,0)
};
\addplot [blue, very thick, dotted, mark=none] coordinates {
(-5,0.8097)
(-4,0.8097)
(-3,0.8097)
(-2,0.8097)
(-1,0.8097)
(0,0)
(1,0)
(2,0)
(3,0)
(4,0)
(5,0)
};
\end{axis}
\end{tikzpicture}

\vspace{0.1cm}
\begin{tikzpicture}
\begin{axis}[
name = axis1,
title = {Panathinaikos vs Dinamo Minsk},
title style = {font=\small},
xlabel = Value of $m$ (goal difference),
x label style = {font=\small},
x tick label style={/pgf/number format/1000 sep=},
xtick = data,
ylabel = Change in probability,
y label style = {font=\small},
width = 0.5\textwidth,
height = 0.42\textwidth,
ymajorgrids = true,
xmin = -5.2,
xmax = 5.2,
ymin = -1.05,
ymax = 1.05,
] 
\addplot [red, thick, dashed, mark=o, mark size=2pt, mark options=solid] coordinates {
(-5,-0.4301)
(-4,-0.3442)
(-3,-0.2204)
(-2,-0.1138)
(-1,-0.0356)
(0,0)
(1,0)
(2,0)
(3,0)
(4,0)
(5,0)
};
\addplot [red, very thick, dotted, mark=none] coordinates {
(-5,-0.0634)
(-4,-0.0634)
(-3,-0.0634)
(-2,-0.0634)
(-1,-0.0634)
(0,0)
(1,0)
(2,0)
(3,0)
(4,0)
(5,0)
};
\addplot [blue, thick, dashdotted, mark=asterisk, mark size=2.5pt, mark options={solid,semithick}] coordinates {
(-5,0.1062)
(-4,0.1031)
(-3,0.0693)
(-2,0.0364)
(-1,0.0171)
(0,0)
(1,0)
(2,0)
(3,0)
(4,0)
(5,0)
};
\addplot [blue, very thick, dotted, mark=none] coordinates {
(-5,0.0259)
(-4,0.0259)
(-3,0.0259)
(-2,0.0259)
(-1,0.0259)
(0,0)
(1,0)
(2,0)
(3,0)
(4,0)
(5,0)
};
\end{axis}

\begin{axis}[
at = {(axis1.south east)},
xshift = 0.08\textwidth,
title = {Rapid Wien vs Copenhagen},
title style = {font=\small},
xlabel = Value of $m$ (goal difference),
x label style = {font=\small},
x tick label style={/pgf/number format/1000 sep=},
xtick = data,
width = 0.5\textwidth,
height = 0.42\textwidth,
ymajorgrids = true,
xmin = -5.2,
xmax = 5.2,
ymin = -1.05,
ymax = 1.05,
] 
\addplot [red, thick, dashed, mark=o, mark size=2pt, mark options=solid] coordinates {
(-5,-0.0098)
(-4,-0.0098)
(-3,-0.0098)
(-2,-0.0098)
(-1,-0.0098)
(0,0)
(1,0.9578)
(2,0.9813)
(3,0.9876)
(4,0.9896)
(5,0.9901)
};
\addplot [red, thick, dotted, mark=none] coordinates {
(-5,-0.0098)
(-4,-0.0098)
(-3,-0.0098)
(-2,-0.0098)
(-1,-0.0098)
(0,0)
(1,0.9704)
(2,0.9704)
(3,0.9704)
(4,0.9704)
(5,0.9704)
};
\addplot [blue, very thick, dashdotted, mark=asterisk, mark size=2.5pt, mark options={solid,semithick}] coordinates {
(-5,0.0998)
(-4,0.0848)
(-3,0.057)
(-2,0.0168)
(-1,0.0098)
(0,0)
(1,0)
(2,0)
(3,0)
(4,0)
(5,0)
};
\addplot [blue, very thick, dotted, mark=none] coordinates {
(-5,0.0261)
(-4,0.0261)
(-3,0.0261)
(-2,0.0261)
(-1,0.0261)
(0,0)
(1,0)
(2,0)
(3,0)
(4,0)
(5,0)
};
\end{axis}
\end{tikzpicture}

\vspace{0.1cm}
\begin{tikzpicture}
\begin{axis}[
name = axis1,
title = {TSC vs Noah},
title style = {font=\small},
xlabel = Value of $m$ (goal difference),
x label style = {font=\small},
x tick label style={/pgf/number format/1000 sep=},
xtick = data,
ylabel = Change in probability,
y label style = {font=\small},
width = 0.5\textwidth,
height = 0.42\textwidth,
ymajorgrids = true,
xmin = -5.2,
xmax = 5.2,
ymin = -1.05,
ymax = 1.05,
] 
\addplot [red, thick, dashed, mark=o, mark size=2pt, mark options=solid] coordinates {
(-5,0)
(-4,0)
(-3,0)
(-2,0)
(-1,0)
(0,0)
(1,0.8255)
(2,0.9244)
(3,0.9508)
(4,0.9874)
(5,0.9947)
};
\addplot [red, very thick, dotted, mark=none] coordinates {
(-5,0)
(-4,0)
(-3,0)
(-2,0)
(-1,0)
(0,0)
(1,0.8918)
(2,0.8918)
(3,0.8918)
(4,0.8918)
(5,0.8918)
};
\addplot [blue, thick, dashdotted, mark=asterisk, mark size=2.5pt, mark options={solid,semithick}] coordinates {
(-5,0.6633)
(-4,0.4896)
(-3,0.3732)
(-2,0.2931)
(-1,0.274)
(0,0)
(1,0)
(2,0)
(3,0)
(4,0)
(5,0)
};
\addplot [blue, very thick, dotted, mark=none] coordinates {
(-5,0.2958)
(-4,0.2958)
(-3,0.2958)
(-2,0.2958)
(-1,0.2958)
(0,0)
(1,0)
(2,0)
(3,0)
(4,0)
(5,0)
};
\end{axis}

\begin{axis}[
at = {(axis1.south east)},
xshift = 0.08\textwidth,
title = {Vit\'oria de Guimar{\~ a}es vs Fiorentina},
title style = {font=\small,align=center},
xlabel = Value of $m$ (goal difference),
x label style = {font=\small},
x tick label style={/pgf/number format/1000 sep=},
xtick = data,
width = 0.5\textwidth,
height = 0.42\textwidth,
ymajorgrids = true,
xmin = -5.2,
xmax = 5.2,
ymin = -1.05,
ymax = 1.05,
legend style = {font=\small,at={(-0.65,-0.25)},anchor=north west,legend columns=2},
legend entries = {Home team$\qquad$,Away team}
] 
\addplot [red, thick, dashed, mark=o, mark size=2pt, mark options=solid] coordinates {
(-5,-0.0366)
(-4,-0.0296)
(-3,-0.00339999999999996)
(-2,-0.000499999999999945)
(-1,-0.000099999999999989)
(0,0)
(1,0)
(2,0)
(3,0)
(4,0)
(5,0)
};
\addplot [blue, thick, dashdotted, mark=asterisk, mark size=2.5pt, mark options={solid,semithick}] coordinates {
(-5,0)
(-4,0)
(-3,0)
(-2,0)
(-1,0)
(0,0)
(1,-0.0366)
(2,-0.0366)
(3,-0.0755)
(4,-0.096)
(5,-0.1363)
};
\addplot [red, very thick, dotted, mark=none] coordinates {
(-5,-0.0029)
(-4,-0.0029)
(-3,-0.0029)
(-2,-0.0029)
(-1,-0.0029)
(0,0)
(1,0)
(2,0)
(3,0)
(4,0)
(5,0)
};
\addplot [blue, thick, dotted, mark=none] coordinates {
(-5,0)
(-4,0)
(-3,0)
(-2,0)
(-1,0)
(0,0)
(1,-0.0447)
(2,-0.0447)
(3,-0.0447)
(4,-0.0447)
(5,-0.0447)
};
\end{axis}
\end{tikzpicture}

\captionsetup{justification=centering}
\caption{Incentives in further matches played in the last \\ round of the 2024/25 UEFA Conference League league phase II. \\ \vspace{0.2cm}
\footnotesize{\emph{Note}: The dotted line shows the (average) change when the value of $m$ is simulated.}}
\label{Fig_A2}
\end{figure}


\end{document}